\documentclass{amsart}

\usepackage{setspace}
\usepackage{hyperref}

\usepackage{amsaddr}
\usepackage[utf8]{inputenc}

\usepackage[margin=1in]{geometry}

\usepackage{amssymb}
\usepackage{amsmath}
\usepackage{amsthm}
\theoremstyle{remark}

\usepackage{graphicx}

\usepackage{wrapfig}
\usepackage{caption}
\usepackage{subcaption}
\usepackage{float}
\usepackage{multirow}

\usepackage{bm}
\usepackage{booktabs}
\usepackage{mathtools}
\usepackage{tikz}
\usepackage{tikz-cd}

\usetikzlibrary{calc}

\usepackage[giveninits,date=year,sortcites]{biblatex}
\usepackage[ruled,vlined]{algorithm2e}

\begin{document}

\title[FEA of Large-Scale Micro-CT Bone Models]{Finite element analysis of very large bone models based on micro-CT scans}

\author[Martinez-Weissberg, Pazner, Yosibash]{Shani Martinez-Weissberg$^1$, Will Pazner$^2$, Zohar Yosibash$^1$}

\address{$^1$School of Mechanical Engineering, The Iby and Aladar Fleischman Faculty of Engineering, Tel-Aviv University, Ramat Aviv, Israel}
\address{$^2$Fariborz Maseeh Department of Mathematics and Statistics, Portland State University, Portland, OR}

\begin{abstract}
High-resolution voxel-based micro-finite element ($\mu$FE) models derived from $\mu$CT imaging enable detailed investigation of bone mechanics but remain computationally challenging at anatomically relevant scales. This study presents a comprehensive $\mu$FE framework for large-scale biomechanical analysis of an intact New Zealand White (NZW) rabbit femur, integrating advanced segmentation, scalable finite element solvers, and experimental validation using predominantly open-source libraries.
Bone geometries were segmented from $\mu$CT data using the MIA clustering algorithm and converted into voxel-based $\mu$FE meshes in Simpleware, which were solved using the open-source MFEM library with algorithms designed for extremely large linear elasticity systems.

The numerical solutions were systematically verified through comparisons with a commercial finite element solver, Abaqus, and by evaluating the performance of full assembly and element-by-element formulations within MFEM. Models containing over $8 \times10^{8}$ degrees of freedom were solved using moderate HPC resources, demonstrating the feasibility of anatomically realistic $\mu$FE simulations at this scale. Resolution effects were investigated by comparing models with voxel sizes of 20, 40, and 80~$\mu$m, revealing that 40~$\mu$m preserves boundary displacement and principal strain distributions with minimal bias while significantly reducing computational cost. Sensitivity analyses further showed that segmentation parameters influence the global mechanical response.

Finally, $\mu$FE predictions were coupled with Digital Image Correlation measurements on an NZW rabbit femur under compression to calibrate effective bone material properties at the micron scale. The results demonstrate that large-scale, experimentally informed $\mu$FE modeling can be achieved using predominantly open-source tools, providing a robust foundation for preclinical assessment of bone mechanics and treatment-related risks.
\end{abstract}

\maketitle

\section{Introduction} \label{s.intro}

The increasing use of small-animal bone models for investigating newly developed treatments requires very large finite element (FE) simulations. For example, laser interstitial thermal therapy (LITT) \cite{Patel2020LaserInterstitialThermalTherapy, Salem2019NeurosurgicalApplication}, used for the ablation of metastatic bone tumors, requires preclinical testing in animal models. The New Zealand White (NZW) rabbit is one of the most widely used models in orthopedic research \cite{Li2015BoneBiomaterials, NeytUseResearch, Harrison2020CorticalOsteoporosis, Harrison2022DirectApproach, Bakici2021ThreeTomography}. To predict the influence of LITT on bone strength, validated animal-specific FE models based on micro-computed tomography ($\mu$CT) are required \cite{Trabelsi2011Patient-specificValidation, Engelke2016FEAReview}. Due to the small size of rabbit bones, $\mu$CT imaging is necessary to accurately capture bone architecture \cite{Bouxsein2010GuidelinesTomography, Clark2014Micro-CTPerspectives}. The mechanical response of rabbit bone is of particular interest because it is the smallest laboratory animal exhibiting cortical bone remodeling behavior comparable to humans \cite{Harrison2020CorticalOsteoporosis, Harrison2022DirectApproach, Pazzaglia2007DesignRabbit, Pazzaglia2009AnatomyFemur}. However, the microscale material properties of rabbit bone remain insufficiently characterized.

To reliably evaluate the biomechanical response of the NZW rabbit's femur given a $\mu$CT scan, a micro-finite element ($\mu$FE$)$ analysis is performed \cite{VanRietbergen1995AModels}, that requires a segmentation process to extract bone tissue from $\mu$CT scans. The choice of segmentation can significantly influence the resulting $\mu$FE model and, consequently, the predicted mechanical response \cite{Bouxsein2010GuidelinesTomography}. Furthermore, the conversion of voxels into hexahedral elements leads to models containing hundreds of millions of elements, which precludes the use of conventional FE solvers.

$\mu$FE models have been investigated extensively in the literature. Oliviero et al.~\cite{Oliviero2021OptimizationApplications}, for example, generated linear $\mu$FE models of the entire mouse tibia based on $\mu$CT scans using a single-threshold segmentation approach. Their models consisted of 8--9 million 8-noded hexahedral elements with homogeneous, isotropic elastic material properties (Young's modulus 14.8~GPa, Poisson's ratio 0.3) and were validated against uniaxial compression experiments. To address memory limitations, a voxel-based $\mu$FE solver was introduced within ParFE \cite{Flaig2011AImages}. Using this approach, linear elastic problems with up to $1.6\times10^{10}$ degrees of freedom (DOFs) were solved on 8000 CPU cores, requiring approximately 1--1.5~TB of distributed memory and achieving solution times of about 12~minutes per linear solve. These large-scale benchmarks were non-anatomical and generated for weak-scaling evaluation by 3D mirroring a single trabecular bone sample embedded in a cubic voxel grid. Stipsitz and Pahr \cite{Stipsitz2018AnModel} demonstrated the feasibility of large-scale nonlinear voxel-based $\mu$FE analysis by extending the ParOSol framework \cite{Flaig2011AImages} and applying it to a $\mu$CT-derived middle section of the human radius comprising approximately 688 million DOFs. The authors demonstrated excellent solver scalability and low memory usage. The analysis, however, was limited to a localized anatomical segment with few details of the model-generation process, and no validation was considered. Recently, Chafia et al.\ \cite{Chafia2024MassivelyComputations} presented a massively parallel finite-element framework for phase-field fracture simulations on distributed-memory supercomputers. Linear systems with up to $10^{10}$ DOFs were solved in a few seconds for idealized structured meshes, while a realistic $\mu$CT-based concrete model comprising 375 million DOFs was simulated using 6000 processes in approximately 69 minutes. The authors reported that simulations with more than $12\times10^{6}$ elements required less than 190 GB of distributed memory, although total memory usage for the largest models was not explicitly reported. Despite these impressive computational achievements, the largest multi-billion-DOF results were limited to simplified geometries, and no experimental mechanical validation was provided.

To the best of our knowledge, no $\mu$FE models with nearly 1 billion DOFs on complex geometries such as a bone have been solved using open-source software. Methods for generating and solving such huge $\mu$FE models of an entire organ have not been thoroughly addressed nor validated by experimental observations.

We herein present a comprehensive methodology for generating very large $\mu$FE models of small animals' long bones, including the segmentation process from $\mu$CT scans, and validating the results by experimental observations. Novel numerical methods to expedite the computational times of the $\mu$FE models, implemented in MFEM \cite{Anderson2021MFEM:Library}, are presented. Sensitivity studies aimed at investigating the segmentation outcome and $\mu$CT resolution on the rabbit's femur mechanical response are addressed. The research's ultimate goal is to determine the Young modulus of the NZW rabbit femurs at the microscale, which could offer valuable insights into bone integrity following cancer treatments.

This paper is structured as follows. In Section \ref{s.MatMethods} we present the generation of a FE model from a $\mu$CT scan, including the segmentation algorithm. Two methods are presented in Section \ref{s.Results} to cope with the huge $\mu$FE model solution. Their advantages and disadvantages are discussed. We present the methodology for calibrating bone material properties at the micron scale by experimental observation using digital image correlation (DIC) in Section \ref{s.Sensitivity} and we conclude with a summary and insights in Section \ref{s.Summary}.

 \section{\texorpdfstring{%
 Generating and Solving the $\mu$FE Model}{%
 Generating and Solving the µFE Model}} \label{s.MatMethods}

The workflow for generating a $\mu$FE model of a rabbit femur from $\mu$CT scans is presented. It consists of three main steps: scanning, segmentation and $\mu$FE model creation.

\subsection{Bone Sample}
The overall algorithm is demonstrated on two NZW rabbbit samples: one from the Roswell Park Comprehensive Cancer Center, PDT Center, Buffalo, NY (denoted L40) and one from the Tel-Aviv School of Medicine (denoted L20), with institutional and legislative approvals. Soft tissue was carefully removed using a scalpel, and the distal end was embedded in a steel base with PMMA for gripping in the loading machine. See Figure~\ref{fig:sample}. When not in preparation, samples were kept frozen at $-20\,^{\circ}\mathrm{C}$, to preserve natural hydration and material properties. The L40 femur had a hole drilled in the lateral shaft for LITT treatment.

\subsection{\texorpdfstring{$\mu$CT Imaging}{µCT Imaging}}
\begin{figure}
            \centering
            \begin{minipage}[t]{0.49\textwidth}
                \centering
                \includegraphics[height=5cm]{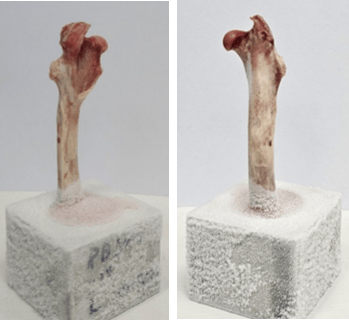}
                \caption{NZW rabbit femur sample preparation. The samples were carefully separated from the rest of the rabbit's leg, soft tissue was cleaned, and the samples were mounted into a steel block using PMMA.}
                \label{fig:sample}
            \end{minipage}
            \hfill
            \begin{minipage}[t]{0.49\textwidth}
                \centering
                \includegraphics[height=5cm]{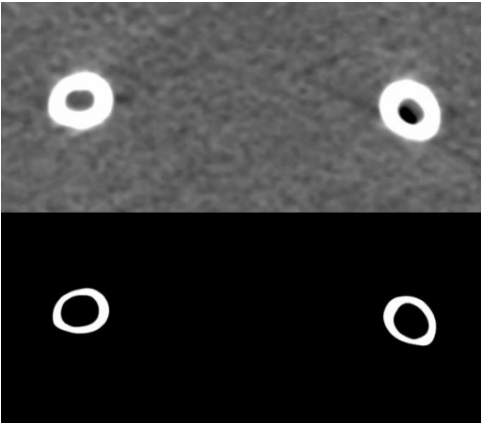}
                \caption{QCT scan of a NZW rabbit femur with slice thickness of 0.67 mm, spacing between slices of 0.33 mm and pixel size of 0.11 × 0.11 $\text{mm}^2$. The same slice of CT scan is presented: top - as received, bottom - after HU thresholding to 750 to match the actual dimensions of the bone. Figure by G. Degabli, 2019 \cite{DegabliGal2019FinalBones}.}
                \label{fig:micro_vs_clinical}
            \end{minipage}
\end{figure}

Since the resolution of clinical quantitative computed tomography (QCT) is not sufficient to properly identify the thin cortical layer of the rabbit femur, these had to be $\mu$CT scanned. Figure \ref{fig:micro_vs_clinical} shows the blurred boundaries of the rabbit's femur as obtained by a clinical CT scan. Mitigating partial-volume effects requires at least two voxels across the thickness of the structure \cite{Bouxsein2010GuidelinesTomography}.

To obtain a sufficient resolution of the cortical region and trabecular structure, a $\mu$CT was used, allowing identification of fine bone geometry \cite{Badea2008InAngiography, Blocker2020TheImaging} and minimizing numerical errors in $\mu$FE models \cite{Guldberg1998TheModels, Niebur1999ConvergenceBone}.

Prior to $\mu$CT scanning, each sample was defrosted and wrapped in a plastic film (e.g., Parafilm) \cite{Synopsys2025SimplewareSoftware} to prevent bone dehydration during the scan \cite{duPlessis2017LaboratorySamples}. The femur samples were scanned using a Nikon XT H 225 ST (Nikon Metrology, NV) with a peak voltage of 185 kV and a current of 88 $\mu$A. The first sample was scanned at 40~$\mu$m resolution, denoted as L40; the second was scanned at 20~$\mu$m resolution, denoted L20. Both scans were reconstructed in Simpleware ScanIP (Version T-2022.03; Synopsys, Inc., Mountain View, USA) from raw TIFF image files exported using the manufacturer's software (Nikon CT Pro 3D), without filtering. As part of the reconstruction, non-relevant background voxels were removed to reduce the output size for downstream processing.

The L40 scan produced 2000 cross-sectional images, capturing the entire bone structure. Image resolution directly affects the number of slices and scan time; consequently, increasing the resolution for the L20 sample reduced the detector's field of view. To capture the full L20 sample, two overlapping scanning windows were used, each producing 2000 cross-sectional images. Reconstruction of the complete L20 volume in Simpleware involved spatial alignment of the two scans based on the known positions of the scanning stage during acquisition and removing irrelevant background areas. This enabled accurate 3D reconstruction of the entire bone at 20~$\mu$m resolution and file size minimization. Both reconstructed scans were saved in DICOM format for the following segmentation process.

\subsection{Segmentation and Creation of the Geometric Bone Model}
Segmentation of $\mu$CT scans is required to create the 3D bone microgeometry by isolating bone tissue voxels from background, soft tissue and air. Its accuracy directly impacts the resulting geometric bone models; therefore, errors can systematically bias subsequent results \cite{Chevalier2007ValidationNanoindentation, Rathnayaka2011EffectsReconstructions}. Despite its importance, a uniform segmentation standard is lacking, and the selection method remains subjective \cite{Bouxsein2010GuidelinesTomography}. A commonly used segmentation approach, valued for its simplicity, applies a global threshold to extract bone voxels exceeding a set value from the $\mu$CT data. However, there is no consistent method for selecting this threshold—some studies rely on visual assessment of the scans, while others use histogram-based techniques \cite{Spoor1993LinearProblems, Fajardo2002AssessingSections, Coleman2007TechnicalModels, Ridler1978CorrespondenceMethod, Ryan2002FemoralPrimates, Fajardo2007NonhumanMode, Dufresne1998SegmentationImaging, Andjelkovic2016VariationTessellata, Cornette2013DoesProblem, DallAra2011AImages, Rajagopalan2005OptimalScaffolds, Meinel2005SilkDefects, Maga2006PreliminaryHominoids, Christiansen2016EffectMice, Oliviero2017EffectTibia}. Despite its popularity, using a single value to segment the whole bone structure can lead to inaccurate representation of structural details.

To improve segmentation accuracy and reproducibility with minimal user input, we employed the open-source Medical Image Analysis (MIA) clustering algorithm developed by Wollny et al.\ \cite{Wollny2013SoftwareAnalysis}. This method classifies each voxel based on the gray values of its local neighborhood and is robust to background noise and gray value inhomogeneity, requiring minimal manual intervention \cite{Dunmore2018MIA-Clustering:Material}. The $\mu$CT DICOM files underwent an automated bone segmentation by MIA\footnote{MIA command: \texttt{mia-3dsegment-local-cmeans -i input.type -o output.type -g XX --cmeans kmeans:nc=Y} where \texttt{-i} and \texttt{-o} specify input/output file types, \texttt{nc} sets the number of clusters \texttt{Y}, \texttt{-g} defines the local grid size in voxels, and \texttt{-t} sets the local threshold (default value of 0.02).} Segmentation parameters were set to 3 clusters (colors) that are locally refined by a 15-voxel grid for L40 and a 25-voxel grid for L20 (after manually evaluating the average trabecular size of 0.4 mm in Simpleware \cite{Synopsys2025SimplewareSoftware}). The grid size was defined to be slightly larger than the largest trabecular thickness to be considered.
The resulting raw files were imported into Simpleware, revealing clear bone voxel segmentation by adjusting the wide range of gray values (GV) to distinguish only three GV levels: bone tissue, soft tissue, and air (background). A binary mask was created that isolates bone voxels, and disconnected voxels were removed.

\subsection{\texorpdfstring{%
$\mu$FE Model}{%
µFE Model}}

\begin{figure}
    \centering
    \begin{subfigure}[b]{0.49\textwidth}
        \includegraphics[width=\linewidth]{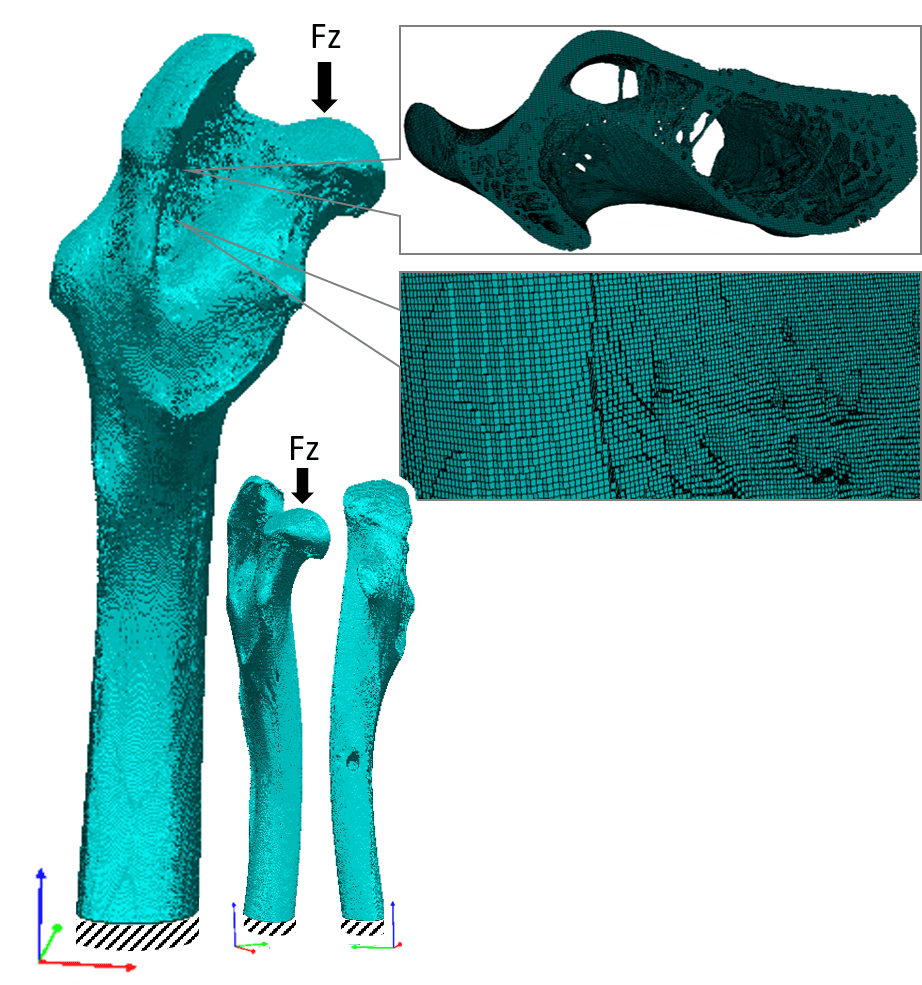}
        \caption{L40 $\mu$FE model.}
        \label{fig:L40MODEL}
    \end{subfigure}
    \hfill
    \begin{subfigure}[b]{0.49\textwidth}
        \includegraphics[width=\linewidth]{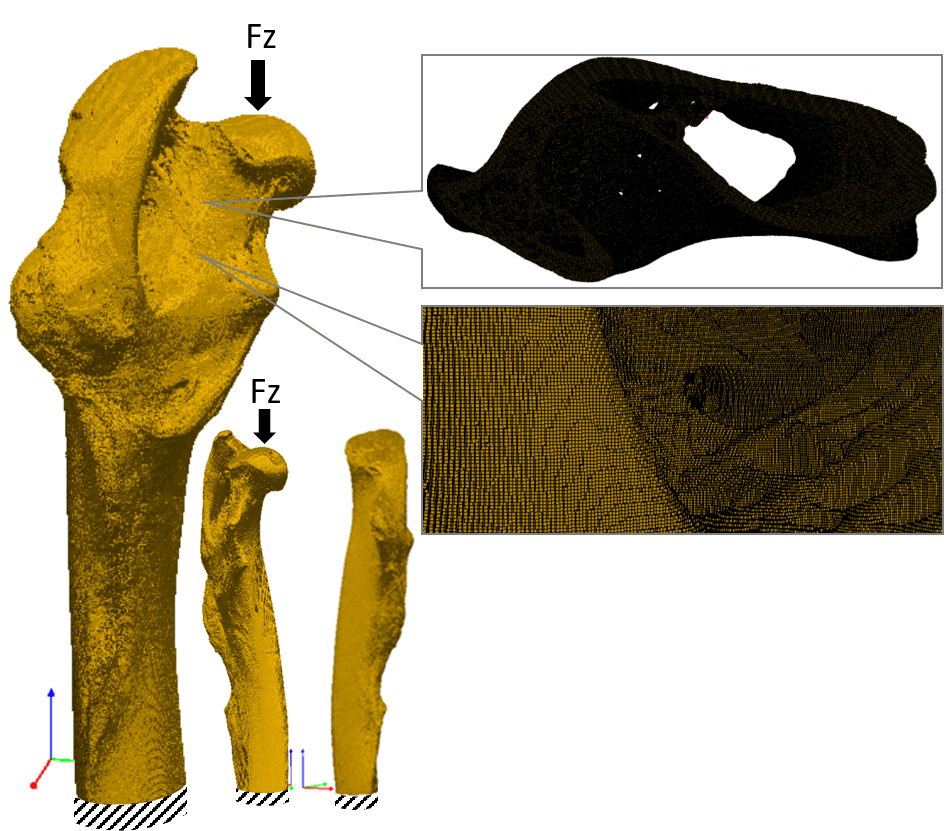}
        \caption{L20 $\mu$FE model.}
        \label{fig:L20MODEL}
    \end{subfigure}
    \caption{Mesh and boundary conditions of the L40 and L20 $\mu$FE femur models. The dashed lines at the distal end represent fully constrained boundary conditions, while the arrows (Fz) indicate the applied compressive traction at the femoral head. Insets illustrate the internal trabecular architecture and voxel-based mesh resolution.}
    \label{fig:20and80FEMesh}
\end{figure}

$\mu$FE models were generated from the segmented $\mu$CT images using Simpleware. Each bone voxel was converted into a linear 8-node hexahedral element, and boundary surfaces were identified. This voxel-based meshing approach produced models with a huge number of degrees of freedom (DOFs), requiring significant computational resources and dedicated solvers (see Fig.\ \ref{fig:20and80FEMesh}).

The MFEM library \cite{Andrej2024High-performanceMFEM, Anderson2021MFEM:Library}, designed for high-performance computing (HPC), was used to solve the $\mu$FE models. A custom script converted the Simpleware mesh output to the supported MFEM \texttt{.mesh v1.x} format. MFEM's parallel linear elasticity solver solved the $\mu$FE model. The bone was considered as an isotropic linear elastic material with an assumed Young modulus of 10 GPa and a Poisson ratio of 0.3. The boundary conditions replicated the experimental setup with a vertical load applied to the femoral head, and the distal end of the shaft clamped (zero displacements). A special $\mu$FE solution technique (detailed in the sequel) was developed for the FE solution. Due to the extensive size of the $\mu$FE model, the visualization of the model results had to be specifically addressed. A custom post-processing algorithm was generated to compute the principal strains and extract the results only on the bone external boundary surface so as to allow comparison with experimental observations, see \ref{postprocess}. This post-processing step enables local visualization of the entire domain, facilitating validation by experimental data.

\begin{figure}
    \centering
    \includegraphics[width=\textwidth]{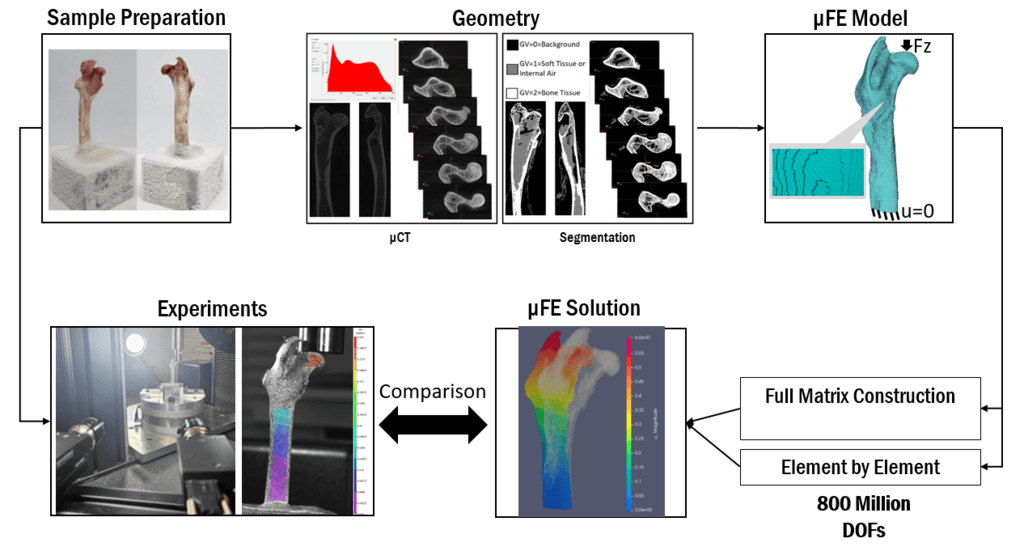}
    \caption{Schematic ﬂow-chart describing the generation of the $\mu$FE model from $\mu$CT scans of a bone.}
    \label{fig:flow-work}
\end{figure}

Overall, the schematic workflow for generating the $\mu$FE model from $\mu$CT scans is presented in Fig.\ \ref{fig:flow-work}.

\subsection{\texorpdfstring{%
MFEM Algorithms for the Linear Elastic Solution of Extreme-Scale $\mu$FE Models}{%
MFEM Algorithms for the Linear Elastic Solution of Extreme-Scale µFE Models}}
\label{s.MFEM_Algorithms}

\newcommand{\T}{\mathcal{T}}
\DeclarePairedDelimiter\ceil{\lceil}{\rceil}
\DeclarePairedDelimiter\floor{\lfloor}{\rfloor}

The solution of such large-scale problems presents algorithmic and discretization-related challenges that have to be addressed. At the same time, the special structure of voxel-based models presents opportunities for optimization and improvement.
The main challenge concerned the solution of the large sparse linear system of equations
\[
   A \bm x = \bm b,
\]
where $A$ is the symmetric positive-definite finite element linear elasticity stiffness matrix.
The size of $A$ is equal to the number of degrees of freedom of the finite element space, and, on voxel-structured meshes, each row of $A$ has, on average, 81 nonzeros (given a 27 point stencil and 3 solution components).

At scale, the solution of this system by means of direct methods such as matrix factorizations is infeasible.
Instead, iterative methods such as conjugate gradient (CG) are typically used.
To achieve scalable performance with the CG method, efficient application of the linear operator $A$ is required, together with a preconditioner $B$, spectrally equivalent to $A^{-1}$, to ensure convergence of the solver in a number of iterations that is independent of the problem size.

In the cases considered presently, there are on the order of $10^9$ degrees of freedom, and so the number of nonzeros in the system matrix is roughly $10^{11}$, requiring approximately 1 TB of memory.
Furthermore, many finite element implementations are memory-bound, meaning that the cost of accessing memory outweighs the cost of performing arithmetic operations \cite{Kolev2021EfficientMethods}.
Consequently, matrix-free methods, which compute the \textit{action} of $A$ without explicitly forming the matrix, promise greater efficiency and better performance \cite{Ljungkvist2014Matrix-freeUnits}.
However, without access to the assembled stiffness matrix, matrix-based preconditioners such as algebraic multigrid cannot be readily used.
This has prompted the development of matrix-free preconditioners, which provide the action of an operator equivalent to $A^{-1}$ without access to the matrix $A$ \cite{Franco2020High-orderPreconditioners,Pazner2020EfficientMethodsb,Pazner2024Matrix,Pazner2023End-to-endDiscretizations}.

To address these challenges, we describe a matrix-free algorithm for applying the elasticity operator and a semi-structured voxel-based coarsening algorithm for constructing a multigrid preconditioner that ensures fast convergence of the iterative solver.

\subsubsection{Matrix-free operator evaluation}

To compute the action of $A$ matrix-free, we decompose $A$ into a nested product of operators (cf.~partial assembly \cite{Andrej2024High-performanceMFEM,Brown2021LibCEED:Discretizations}),
\begin{equation}
   \label{eq:decomposition}
   A = P^T G^T \widehat{A} G P,
\end{equation}
where $P$ is a \textit{parallel prolongation matrix}, that represents parallel communication between mesh partitions distributed across MPI ranks, and $G$ represents \textit{finite element assembly}, encoding the mapping between element-local degrees of freedom and global degrees of freedom;
the matrix $\widehat{A}$ is block-diagonal, where each diagonal block is the element-local stiffness matrix.
The action of $P$ and $G$ (and their transposes) can be computed matrix-free using connectivity tables.
In the case of voxel-structured meshes considered presently (with constant or piecewise-constant coefficients), all blocks of $\widehat{A}$ are identical (or multiples of each other), and so the storage requirements are greatly reduced.
Furthermore, the action of $\widehat{A}$ can be computed very efficiently using batched linear algebra routines.

\subsubsection{Voxel-structured multigrid hierarchy}

We consider the use of multigrid as a preconditioner for the conjugate gradient method.
Multigrid methods have the important property that, for a large class of problems, the condition number of the resulting preconditioned system remains bounded, independent of the problem size.
Furthermore, the cost of applying the multigrid preconditioner scales linearly with the problem size, resulting in a solver with optimal complexity ($\mathcal{O}(N)$, where $N$ is the number of degrees of freedom).
Standard multigrid methods assume that there exists a nested hierarchy of meshes $\T_0, T_1, \ldots, T_J := \T$, resulting in a nested sequence of finite element spaces,
\[
   \begin{tikzcd}
   V_0 \arrow[r, hook, "P_1"]
   & V_1 \arrow[r, hook, "P_2"]
& \cdots \arrow[r, hook, "P_J"]
   & V_J := V_h
   \end{tikzcd}
\]
where the \textit{prolongation operators} $P_i : V_{i-1} \to V_i$ are given by the natural inclusion.
In this context, each mesh $\T_i$ is usually obtained from the coarser mesh $\T_{i-1}$ through a refinement procedure.

On large-scale voxel meshes obtained from $\mu$CT scans, a hierarchy of nested meshes is not available.
The voxel mesh $\T$ is not obtained from coarser meshes through refinement, and $\T$ possess complicated topological features and small-scale geometries that make coarsening or derefinement challenging.
To address this issue, we consider the generation of a sequence of coarse meshes $\T_i$, starting from the original mesh $\T$ through a semi-structured voxel coarsening algorithm.
In contrast to standard multigrid methods, the domain $\Omega_i$ of the coarsened mesh $\T_i$ may not coincide with the problem domain $\Omega$.
Instead, we require that $\Omega_{i+1} \subseteq \Omega_i$.
The simple voxel-structured coarsening algorithm is described in Algorithm \ref{alg:coarsening}.

\begin{algorithm}
   \caption{Semi-structured voxel coarsening}
   \label{alg:coarsening}

   \AlgoDisplayBlockMarkers\SetAlgoBlockMarkers{}{end}\SetAlgoNoEnd
   \DontPrintSemicolon
   \setstretch{1.1}

   \SetKwFunction{EmptyGrid}{EmptyGrid}
   \SetKwFunction{IsMarked}{IsMarked}
   \SetKwFunction{Mark}{Mark}

   \KwIn{Fine mesh $\T_{\mathrm{fine}}$}
   \KwOut{Coarse mesh $\T_{\mathrm{coarse}}$}
   Let $m_x, m_y, m_z$ denote the grid dimensions of $\T_{\mathrm{fine}}$ \;
   $n_x \gets \ceil{m_x/2}$, \quad $n_y \gets \ceil{m_y/2}$, \quad $n_z \gets \ceil{m_z/2}$ \;
   $\T_{\mathrm{coarse}} \gets $ \EmptyGrid{$n_x, n_y, n_z$} \;
   \For{$(i,j,k)$ in $[0,m_x) \times [0,m_y) \times [0,m_z)$}{
      \If {\IsMarked{$\T_{\mathrm{fine}}, i, j, k$}}{
         $i' \gets \floor{i/2}$, \quad $j' \gets \floor{j/2}$, \quad $k' \gets \floor{k/2}$ \;
         \Mark{$\T_{\mathrm{coarse}}, i, j, k$} \;
      }
   }
   \Return{$\T_{\mathrm{coarse}}$} \;
\end{algorithm}

The result of this coarsening procedure is a sequence of meshes $\T_0, \T_1, \ldots, T_J = \T$, which are not strictly nested through refinements.
Therefore, the associated finite element spaces $V_0, V_1, \ldots, V_J = V_h$ are not nested.
However, it is still possible to define natural prolongation operators $P_i : V_{i-1} \to V_i$ as follows.
Given $v_{i-1} \in V_{i-1}$, $P v_{i-1} \in V_i$ is defined by first mapping $v_{i-1} \mapsto \tilde{v}_i$, in a uniformly refined space on the potentially enlarged domain $\Omega_{i-1}$.
Then, $v_i \in V_i$ is obtained from $\tilde{v}_i$ by restricting to $\Omega_i$ and applying essential boundary conditions at Dirichlet boundaries.

\subsubsection{Smoothers and Chebyshev acceleration}

At each level of the multigrid hierarchy, we applied a smoother.
The voxel hierarchy allows for matrix-free computation of the action of the operator;
Consequently, matrix-based smoothers such as Gauss--Seidel and incomplete factorizations are infeasible.
Presently, we considered the use of block Jacobi smoothers, and their variants accelerated using Chebyshev polynomials \cite{Adams2003ParallelGaussSeidel,Golub1961ChebyshevMethods}.

The block Jacobi smoother is given by the action of $D^{-1}$, where $D$ is the block-diagonal part of the matrix $A$ (with blocks of size $d \times d$).
Each block corresponds to all the degrees of freedom belonging to a given physical node in the finite element space.
Note that the block-diagonal part of $A$ can be computed matrix-free as follows.
Let $\widehat{A}$ denote the elementwise broken operator in the decomposition \eqref{eq:decomposition}.
As noted previously, in the case of voxel meshes, all blocks of $\widehat{A}$ are identical, and only one element matrix needs to be formed.
Let $\widehat{D}$ correspond to the $d \times d$ diagonal blocks of $\widehat{A}$.
Then, $D = P^T G^T \widehat{D}$ gives the block-diagonal part of $A$, as can be seen as follows.

Let $\Lambda = GP$.
This is a boolean matrix, mapping global unique degrees of freedom to elementwise broken degrees of freedom.
Each broken DOF corresponds to one global DOF, and hence each row of $\Lambda$ has exactly one nonzero entry.
The nonzeros of each column are the broken (duplicated DOFs) corresponding to a global DOF.
All nonzeros in a given column correspond to the broken DOFs lying in distinct elements, which implies that, for any $i$, if $\Lambda_{ki} \Lambda_{\ell i} \neq 0$, then $k$ and $\ell$ correspond to distinct elements, and so $\widehat{A}_{k\ell} = 0$, since $\widehat{A}$ is elementwise block-diagonal.

We then compute the (block-) diagonal entries of $A = \Lambda^T \widehat{A} \Lambda$,
\begin{align*}
   A_{ii} &= (\Lambda^T \widehat{A} \Lambda)_{ii}
      = \sum_\ell \Lambda_{\ell i} \sum_k \widehat{A}_{\ell k} \Lambda_{ki}
      = \sum_{\ell k} \Lambda_{\ell i} \widehat{A}_{\ell k} \Lambda_{ki} \\
      &= \sum_k \Lambda_{k i} \widehat{A}_{kk} \quad\text{since $\Lambda_{ki}\Lambda_{\ell i} \neq 0$ for $k \neq \ell$ implies $\widehat{A}_{\ell k} = 0$,} \\
      &= (\Lambda_T \widehat{D})_{ii} = D_{ii}.
\end{align*}

The block Jacobi smoother can be improved through the use of Chebyshev acceleration.
In this context, a sequence of block Jacobi iterates are generated through the recurrence,
\begin{align*}
   \bm x^{(0)} &= 0, \\
   \bm x^{(i+1)} &= \bm x^{(i)} + D^{-1}(\bm b - A \bm x^{(i)}).
\end{align*}
The $m$th fixed-point iterate is given by
\begin{align*}
   \bm x^{(m)} &= \sum_{j=0}^{m} (I - D^{-1} A)^j D^{-1} \bm b \\
      &= p(D^{-1} A) D^{-1} \bm b,
\end{align*}
where the degree-$m$ polynomial $p$ is defined by $p(t) = \sum_{j=0}^{m} (1 - t)^j$.
The error associated with this fixed-point iteration is given by
\begin{align*}
   \bm e^{(m)} &= \bm x - \bm x^{(m)} \\
      &= \bm x - p(D^{-1} A) D^{-1} \bm b \\
      &= \bm x - p(D^{-1} A) D^{-1} A \bm x \\
      &= (I - p(D^{-1}A) D^{-1}A) \bm e^{(0)} \\
      &= q(D^{-1} A) \bm e^{(0)}.
\end{align*}
The error reduction is characterized by the polynomial $q(t) := 1 - p(t)t = (1-t)^{m+1}$.
A more effective smoother can be obtained by choosing a different polynomial $\tilde{q}(t)$ (which will correspond to a different iteration polynomial $\tilde{p}$) to damp the high-frequency errors.
If the high-frequency part of the spectrum of $D^{-1}A$ is denoted $[\lambda_*, \lambda_{\max}]$, then $\tilde{q}(t)$ can be chosen such that $\max_{t \in [\lambda_*, \lambda_{\max}]} |q(t)|$ is minimized, under the constraint that $\tilde{q}(0) = 1$;
the solution to the minimax problem is given by (shifted) Chebyshev polynomials of the first kind, i.e.
\[
   \tilde{q}(t) := \frac{T_m\left( (t - \theta) / \delta \right)}{T_m\left( -\theta / \delta \right)}, \quad
   \theta = \frac{\lambda_{\max} + \lambda_*}{2}, \quad
   \delta = \frac{\lambda_{\max} - \lambda_*}{2}.
\]
The corresponding Chebyshev iteration is given by
\[
   \bm x^{(m)}_C = \tilde{p}(D^{-1}A) D^{-1} \bm b,
\]
where $\tilde{p}(t) = t^{-1}(1 - \tilde{q}(t))$.

\subsubsection{Performance comparison}

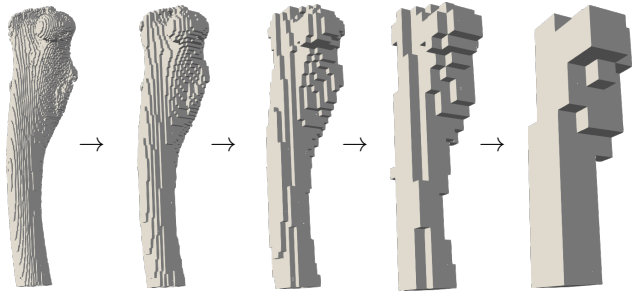
\begin{figure}
   \centering
   \begin{tikzpicture}
      \node [rotate=90] (A1) {\includegraphics[scale=0.17]{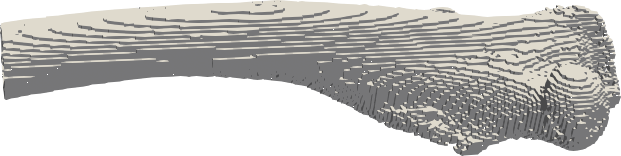}};
      \node [rotate=90] (A2) at ($(A1) + (1.75,0.0)$) {\includegraphics[scale=0.17]{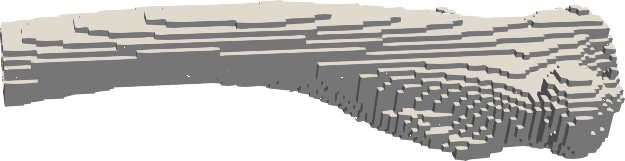}};
      \node [rotate=90] (A3) at ($(A2) + (1.75,0.0)$) {\includegraphics[scale=0.17]{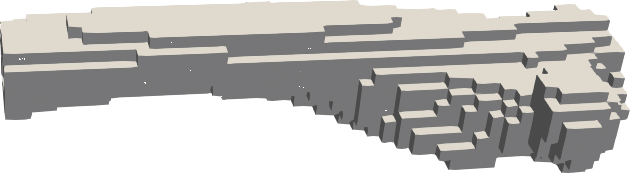}};
      \node [rotate=90] (A4) at ($(A3) + (1.75,0.0)$) {\includegraphics[scale=0.17]{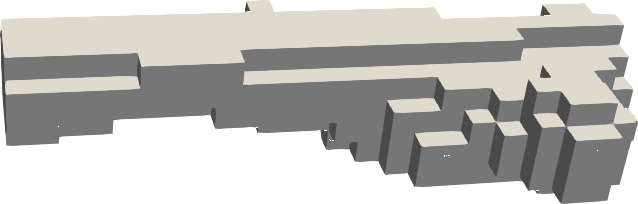}};
      \node [rotate=90] (A5) at ($(A4) + (1.9,0.0)$) {\includegraphics[scale=0.16]{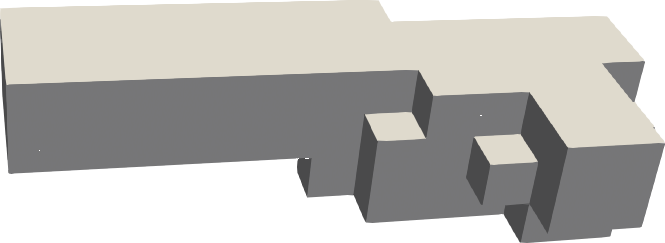}};

      \node [xshift=-6pt] at ($(A1)!0.5!(A2)$) {$\to$};
      \node [xshift=-6pt] at ($(A2)!0.5!(A3)$) {$\to$};
      \node [xshift=-6pt] at ($(A3)!0.5!(A4)$) {$\to$};
      \node [xshift=-6pt] at ($(A4)!0.5!(A5)$) {$\to$};

      \coordinate (M) at ($(A1)!0.5!(A4)$);
      \coordinate (T) at (0,-2.1);
   \end{tikzpicture}
   \caption{Reduced femur model with 72,070 hexahedral elements and associated coarsened voxel hierarchy.}
   \label{fig:voxel-mg}
\end{figure}

In this section, we compare the performance, in terms of runtime, number of iterations, and memory usage, of the matrix-free voxel-based multigrid solver, compared with black-box algebraic multigrid.
As a test problem, we consider a reduced femur voxel model, consisting of 72,070 hexahedral elements.
This mesh and voxel-coarsened versions of it are shown in Figure \ref{fig:voxel-mg}.
We then perform $\ell$ levels of uniform refinement to obtain a finer mesh.
A constant-coefficient linear elasticity problem was then solved.
For this test case, we used pure displacement boundary conditions.
In the case of mixed or traction boundary conditions, the matrix-free voxel-based coarsening algorithm exhibits somewhat degraded convergence; addressing this difficulty is the subject of future work.

A comparison of memory usage and runtime for the voxel-based multigrid (VMG) algorithm and algebraic multigrid (AMG) is shown in Table \ref{tab:voxel-mg}.
The number of conjugate gradient iterations is roughly comparable between the two methods.
The matrix-free voxel multigrid demonstrated slightly better robustness under uniform mesh refinement.
The most significant difference between the two methods is in the assembly time.
Since the matrix-free method requires assembly of only a single element matrix, the assembly time on the finest mesh ($\ell = 3$) was reduced from approximately 300 seconds to $3 \times 10^{-4}$ seconds.
The multigrid setup time was consistently faster for the voxel multigrid than algebraic multigrid; the speed-up ranges from 2.5$\times$ for $\ell = 1$ to 4.5$\times$ for $\ell = 3$.
The CG solve time was roughly comparable for most cases, though with increasing refinement, VMG performed more favorably than AMG.
For $\ell = 3$, the CG solve with the voxel method was about 1.7$\times$ faster than with AMG.
Finally, for large problems, the memory requirements are significantly reduced using the matrix-free method.
This is because, as discussed above, the memory cost associated with storing the finest-level system matrix is quite large.
In contrast, the matrix-free method requires storing only a single element matrix together with the mesh topology information required to construct the finite element space.
At the finest level, the matrix-free method required about 3.4$\times$ less memory than the matrix-based method.

\begin{table}
   \caption{Solver performance for matrix-free voxel multigrid (VMG) and algebraic multigrid (AMG) on the reduced femur test case.}
   \label{tab:voxel-mg}

   \vspace{\baselineskip}
   \centering

   \begin{tabular}{cc|cclllc}
      \toprule
      $\ell$ & \# DOFs & Method & CG It. & Assembly (s) & Setup (s) & Solve (s) & Memory (GB) \\
      \midrule
      \multirow{2}{1em}{0} &
      \multirow{2}{5em}{\centering $3.34\times 10^{5}$} &
      AMG & 6 & $1.73\times 10^{-1}$ & $4.68\times 10^{-2}$ & $6.15\times 10^{-2}$ & 23.9 \\
      &  & VMG & 6 & $1.07\times 10^{-4}$ & $3.39\times 10^{-2}$ & $6.52\times 10^{-2}$ & 23.4 \\
      \midrule
      \multirow{2}{1em}{1} &
      \multirow{2}{5em}{\centering $2.22\times 10^{6}$} &
      AMG & 7 & $1.56\times 10^{0}$ & $4.54\times 10^{-1}$ & $8.29\times 10^{-1}$ & 29.0 \\
      &  & VMG & 6 & $2.09\times 10^{-4}$ & $1.82\times 10^{-1}$ & $6.29\times 10^{-1}$ & 24.7 \\
      \midrule
      \multirow{2}{1em}{2} &
      \multirow{2}{5em}{\centering $1.58\times 10^{7}$} &
      AMG & 7 & $1.66\times 10^{1}$ & $4.54\times 10^{0}$ & $8.24\times 10^{0}$ & 66.0 \\
      &  & VMG & 6 & $1.14\times 10^{-4}$ & $1.55\times 10^{0}$ & $6.91\times 10^{0}$ & 33.9 \\
      \midrule
      \multirow{2}{1em}{3} &
      \multirow{2}{5em}{\centering $1.19\times 10^{8}$} &
      AMG & 8 & $2.98\times 10^{2}$ & $5.91\times 10^{1}$ & $9.62\times 10^{1}$ & 354.6 \\
      &  & VMG & 6 & $3.48\times 10^{-4}$ & $1.31\times 10^{1}$ & $5.58\times 10^{1}$ & 105.4 \\
      \bottomrule
   \end{tabular}
\end{table}
 \section{\texorpdfstring{%
 $\mu$FE Solutions and their Verification}{%
 µFE Solutions and their Verification}} \label{s.Results}

\begin{figure} \centering
    \includegraphics[width={0.2\textwidth}]{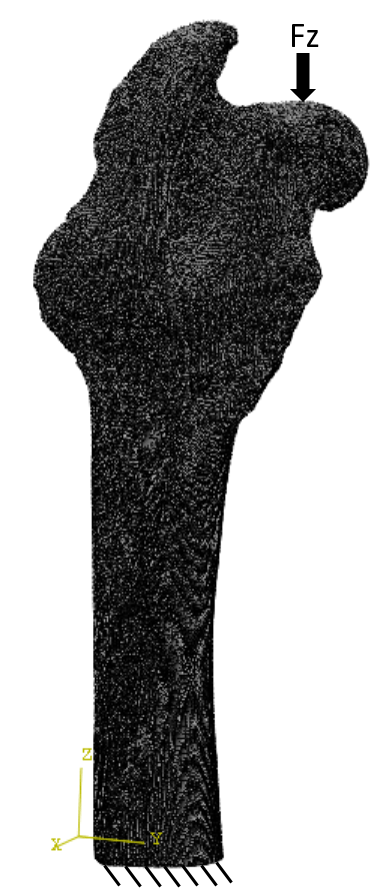}
    \caption{FE L20 coarse model (generated from the L20 $\mu$CT scan and boundary conditions having 3.6M linear hexahedral elements).}
    \label{fig:AbaqusBC}
\end{figure}

As a starting point, we verified the successful implementation of MFEM for the solution of the bone model by comparison with the commercial FEA product Abaqus/CAE (v2022, Dassault Systèmes Simulia Corp., Providence, RI, USA). For this purpose, a coarse FE model has been considered. The L20 $\mu$CT was resampled by merging 4×4×4 original $\mu$CT voxels (20$^3~\mu\text{m}^3$ each) into elements of 80$^3~\mu\text{m}^3$ each. Abaqus and MFEM were used to solve the same coarsened $\mu$FE model, which contained 3,593,570 linear hexahedral elements (4,646,535 nodes). A vertical surface traction was applied to the top surface of the femur head to mimic a total vertical force of 91.4 N, with the bottom surface clamped. Linear isotropic material properties were assigned with a $E=10$ GPa and $\nu$=0.3. Figure~\ref{fig:AbaqusBC} illustrates the FE mesh with boundary conditions. The resampled model contains almost 14M DOFs and 12 processors were used to solve the model by both applications.

\begin{figure}
    \centering
    \begin{subfigure}[b]{0.49\textwidth}
        \includegraphics[height=7cm]{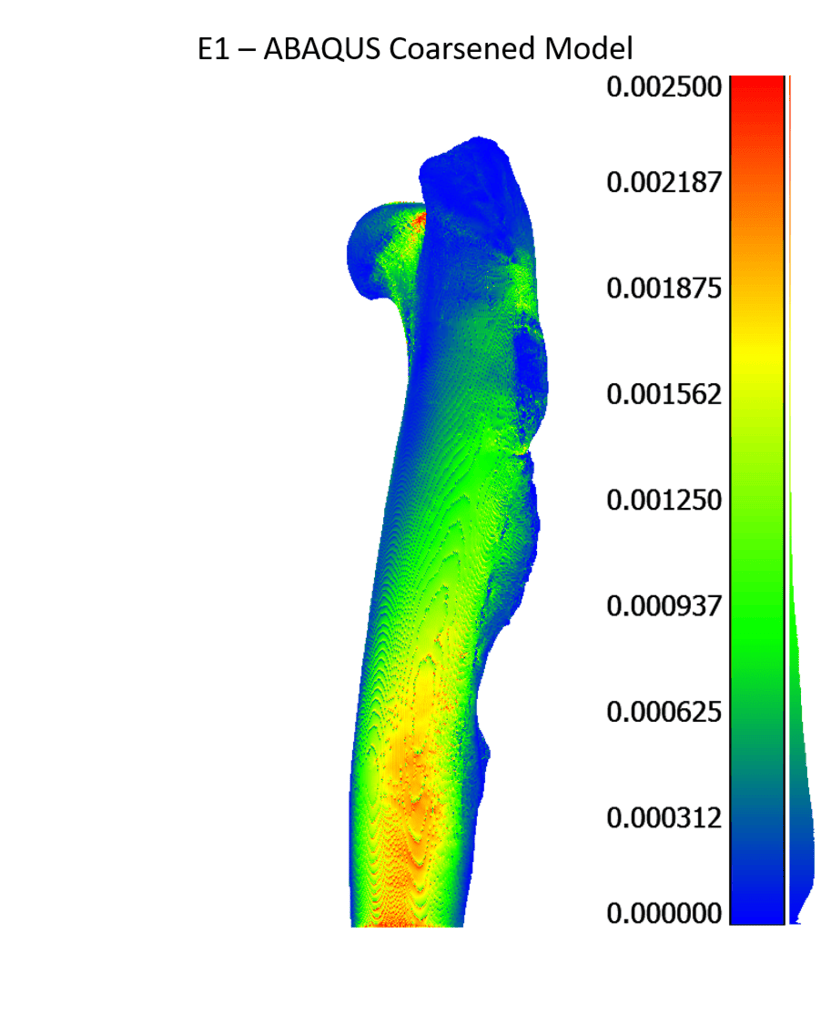}
        \caption{Abaqus E1}
        \label{fig:AbaqusE1}
    \end{subfigure}
    \hfill
    \begin{subfigure}[b]{0.49\textwidth}
        \includegraphics[height=7cm]{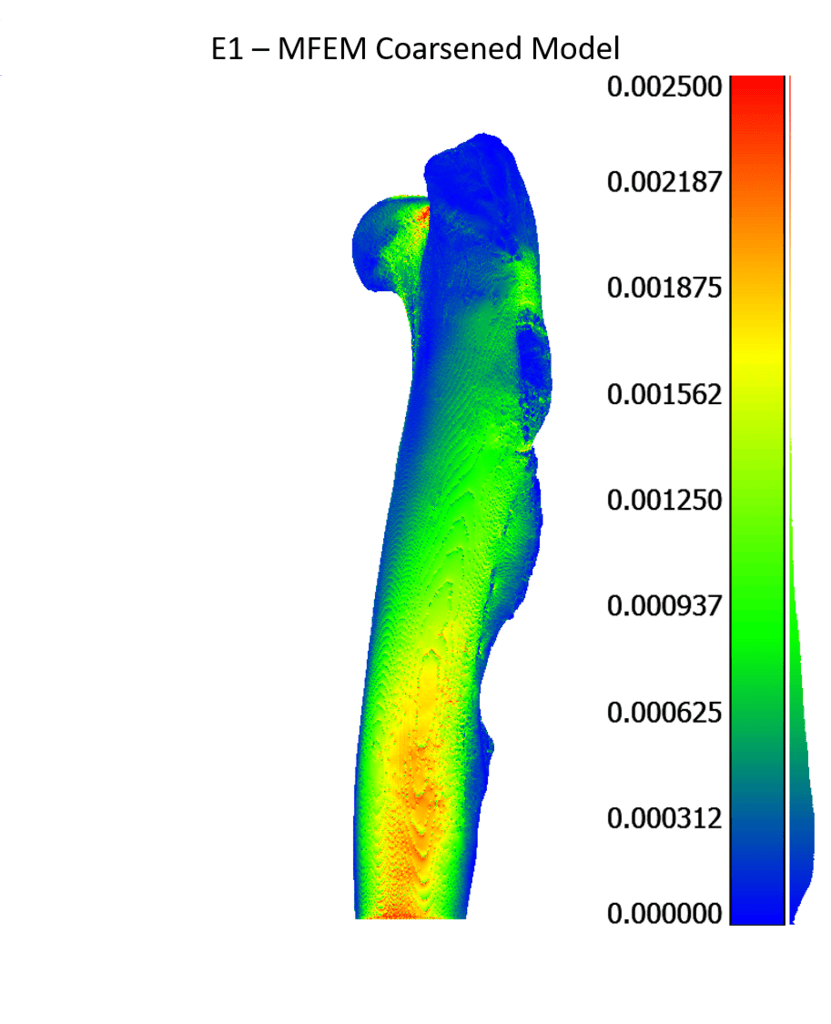}
        \caption{MFEM E1}
        \label{fig:mfemE1}
    \end{subfigure}
    \caption{Comparison between Abaqus and MFEM maximum principal strain field, E1, of the L20 coarsened FE model.}
    \label{fig:E1mfem-vs-abaqus}
\end{figure}

\begin{figure}
    \centering
    \begin{subfigure}[b]{0.49\textwidth}
        \includegraphics[height=7cm]{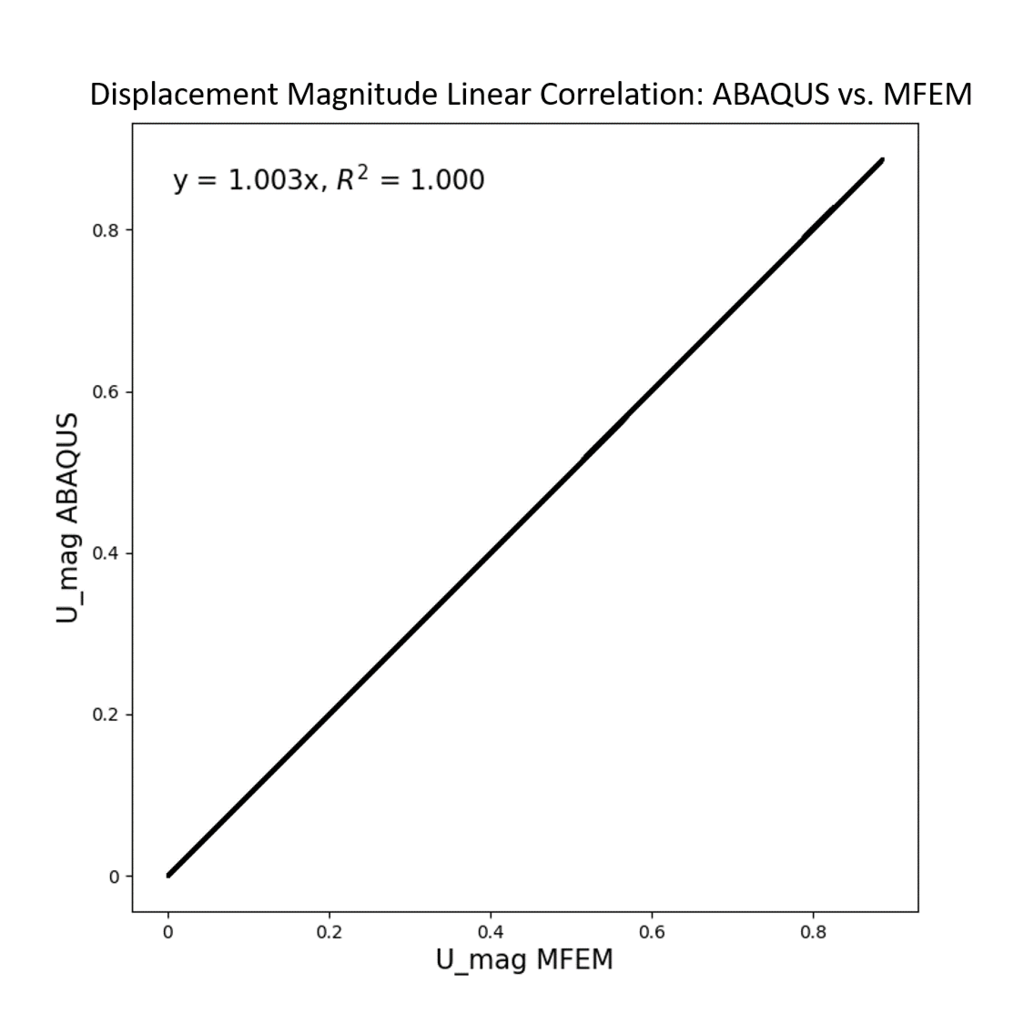}
        \caption{Displacement magnitude}
        \label{fig:umag}
    \end{subfigure}
    \hfill
    \begin{subfigure}[b]{0.49\textwidth}
        \includegraphics[height=7cm]{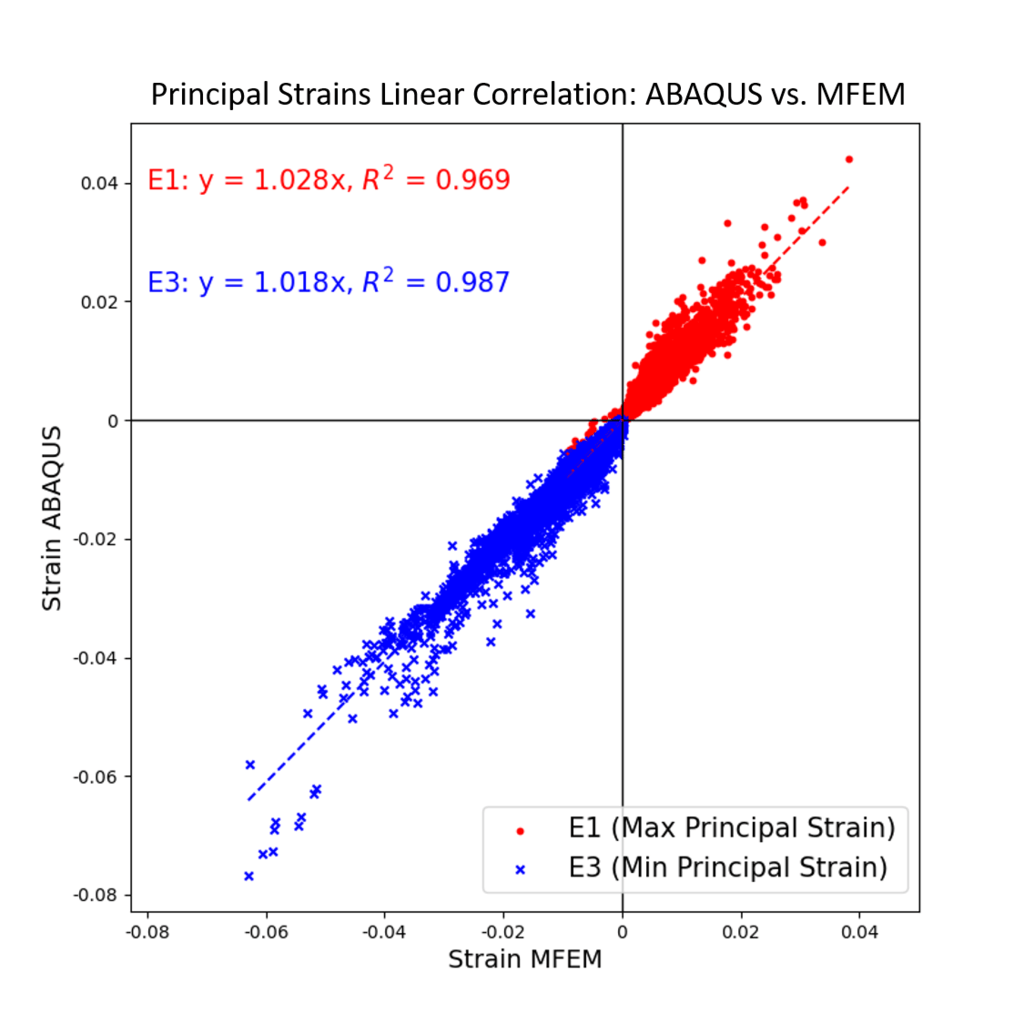}
        \caption{Principal strains E1 and E3}
        \label{fig:e1e3}
    \end{subfigure}
    \caption{Comparison between Abaqus and MFEM results of the L20 coarsened FE model. 4,646,535 neighbor data points were compared.}
    \label{fig:mfem-vs-abaqus}
\end{figure}

The displacements and principal strains at all nodes (4,646,535 points) were compared. Fig.\ \ref{fig:E1mfem-vs-abaqus} shows the graphical representation of the maximum principal strains obtained by the two codes. A perfect agreement between the solutions was observed, as shown in the linear correlation between the two displacement magnitudes $(U_{mag})$ in Fig.\ \ref{fig:umag}. Because of the different algorithms for evaluating the discontinuous strain fields at the nodes in the two FE codes, there are minor discrepancies between the principal strains as shown in Fig.\ \ref{fig:e1e3}.

The verification outcome demonstrates the proper usage of MFEM code, allowing us to proceed to the huge FE model.

\subsection{\texorpdfstring{%
Solution of Very Large-Scale $\mu$FE Models
}{%
Solution of Very Large-Scale µFE Models
}}
\label{s.MFEM}

The two huge $\mu$FE femur models, L40 and L20, are solved by the two MFEM algorithms presented in Section \ref{s.MFEM_Algorithms}. The first uses the global stiffness matrix stored in memory, denoted as ``Full Assembly (FA)''. The second is denoted ``Element by Element (EBE)'', without forming the global stiffness matrix. Both algorithms are applied to the two $\mu$FE models, the L20 and L40. The total memory usage (the sum of the peak memory used by the process on each node), total run time, and number of linear solver iterations are presented in Table ~\ref{tab:comparison}.  Both algorithms used 5 nodes with a total of 192 MPI ranks (CPU cores). For the L20 model the EBE method allocated 75\% less memory but 63\% more time than the FA method.
\begin{table}[ht]
    \begin{center}
    \caption{Memory usage, time, and residuals for the different algorithms.}
    \label{tab:comparison}
    \begin{tabular}{lccccccc}
        \toprule
        Mesh & \# of & DOFs & Algorithm & Memory & Time & \# of & Residual \\
          &  Elements &      &       & [GB]   & [min] & iterations  &   \\
        \midrule
        \multirow{2}{*}{L20} & \multirow{2}{*}{251,302,209} & \multirow{2}{*}{811,375,038} & FA & 2818 & 245.4 & 840 & \( r = 9.70 \times 10^{-9} \) \\
         &  &  & EBE & 692 & 399.0 & 2218 & \( r^2 = 4.00 \times 10^{-17} \) \\
        \midrule
        \multirow{2}{*}{L40} & \multirow{2}{*}{39,325,803} & \multirow{2}{*}{132,619,929} & FA & 499 & 11.4 & 302 & \( r = 9.87 \times 10^{-9} \) \\
         &  &  & EBE & 159 & 17.5 & 709 & \( r^2 = 1.4 \times 10^{-18} \) \\
        \bottomrule
    \end{tabular}
\end{center}
\end{table}

\subsection{Post-Processing}\label{postprocess}
A custom post-processing algorithm was developed to compute the principal strains and extract the boundary data (mesh, displacements, and principal strains) from the large output files of the $\mu$FE models. The MFEM SubMesh interface allows efficient extraction of flagged boundary surfaces and robust transfer of grid functions (displacements and principal strains) between the parent mesh and its boundary submesh \cite{Andrej2024High-performanceMFEM}.
The post-processing step presented in algorithm \ref{a.PP} enables local visualization of the entire domain, facilitating validation against experimental data.
\begin{algorithm}
    \caption{Post-processing}
    \label{a.PP}
    \KwIn{The $\mu$FE model partitioned mesh and displacement results in MFEM format}
    \KwOut{The $\mu$FE model boundary partitioned mesh and displacement and strain results in ParaVeiw format}
    Initialize MPI\;
    Load partitioned mesh and displacement data\;
    Define finite element spaces\;
    Compute strain tensor and principal strains\;
        \For{each IP}{
            Compute gradient of displacement\;
            Symmetrize to obtain strain tensor\;
            Calculate eigenvalues (principal strains)\;
        }
    Extract boundary submesh\;
    Interpolate displacement and strain data onto submesh\;
\end{algorithm}

To compare the results between the FA and EBE solvers, the boundary surfaces from the two solutions were rigidly registered in CloudCompare (version 2.13.2) \cite{2024CloudCompareSoftware}. A custom nearest-neighbor search (fixed-radius pairing) ensured that only spatially corresponding points present in both models were included in the correlation analysis. For the L20 $\mu$FE model, all boundary nodes ($\sim$40 M paired points) were analyzed.
The linear correlations
between the results of the FA and EBE exhibit unity slopes \text{1.000}) and negligible intercepts ($\approx 10^{-10}$), with coefficients of determination $R^2=1.00$ for Umag, E1, and E3, indicating perfect numerical agreement between the two algorithms.

\begin{figure}
    \centering \includegraphics[width=0.75\textwidth]{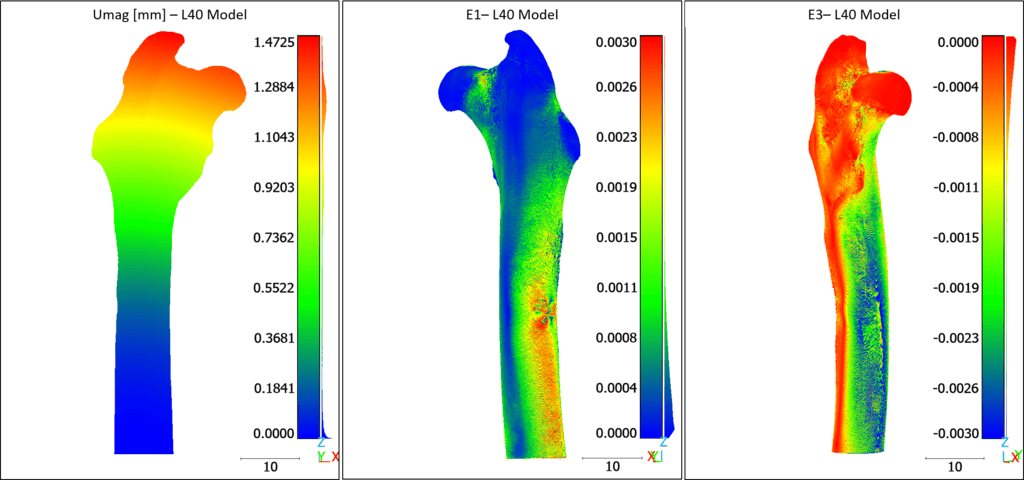}
    \caption{MFEM's L40 model results using the FE and EBE methods. Left: Displacement magnitude (Umag) in mm. Middle: Maximum principal strain (E1), and Right: Minimum principal strain (E3).}
    \label{fig:L40FAvsEBE_result}
\end{figure}

To further verify equivalence, the L40 $\mu$FE model was also solved using both FA and EBE. All boundary nodes ($\sim$10 M paired points) were included. The displacement and strain fields are visually indistinguishable between methods (Figure~\ref{fig:L40FAvsEBE_result}). The corresponding correlation plots
showed again a slope of 1.000 with intercepts on the order of $10^{-16}$--$10^{-14}$ and $R^2=1.00$ across all quantities.

These results confirm that FA and EBE methods produce mathematically identical boundary-level displacements and principal strains for both voxel resolutions (20$\mu$m and 40$\mu$m). No systematic bias or numerical deviation were detected.
\begin{figure}
    \centering \includegraphics[width=0.75\textwidth]{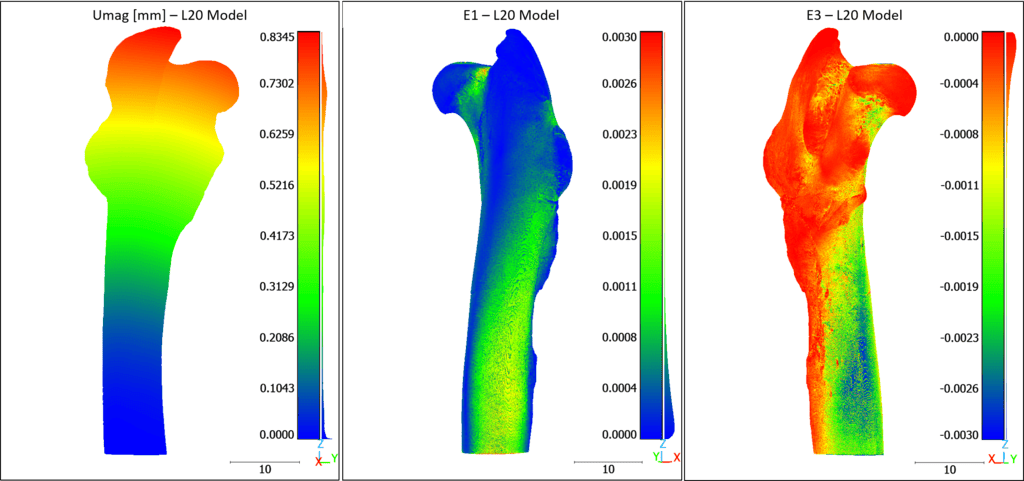}
    \caption{MFEM's L20 model results using the FE and EBE methods. Left: Displacement magnitude (Umag) in mm. Middle: Maximum principal strain (E1), and Right: Minimum principal strain (E3).}
    \label{fig:FAvsEBE_result_Z}
\end{figure}

\subsubsection{Comparison of 20 $\mu$m, 40 $\mu$m, and 80 $\mu$m Models Based on the L20 Femur}\label{80vs40vs20}

\begin{figure}[H]
    \centering
    \begin{subfigure}[b]{0.49\textwidth}
        \includegraphics[width=\linewidth]{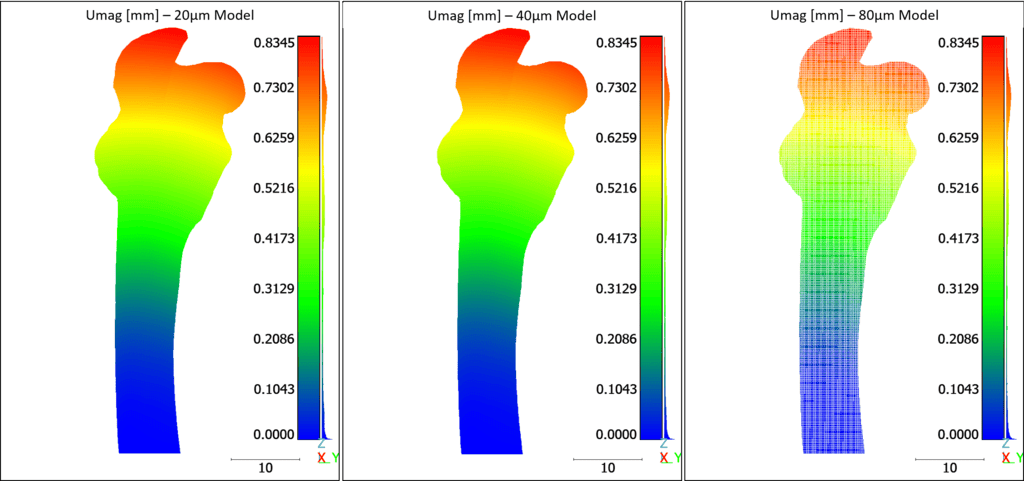}
        \caption{Umag, Displacement magnitude [mm].}
        \label{fig:Umag20vs40vs80}
    \end{subfigure}
    \vspace{1em}

    \begin{subfigure}[b]{0.49\textwidth}
        \includegraphics[width=\linewidth]{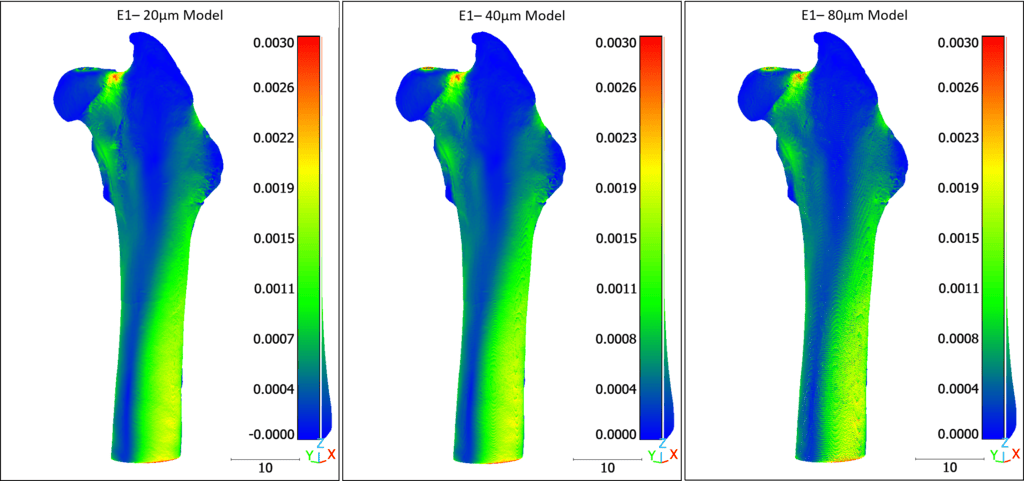}
        \caption{E1, Maximum principal strains.}
        \label{fig:E120vs40vs80}
    \end{subfigure}
    \hfill
    \begin{subfigure}[b]{0.49\textwidth}
        \includegraphics[width=\linewidth]{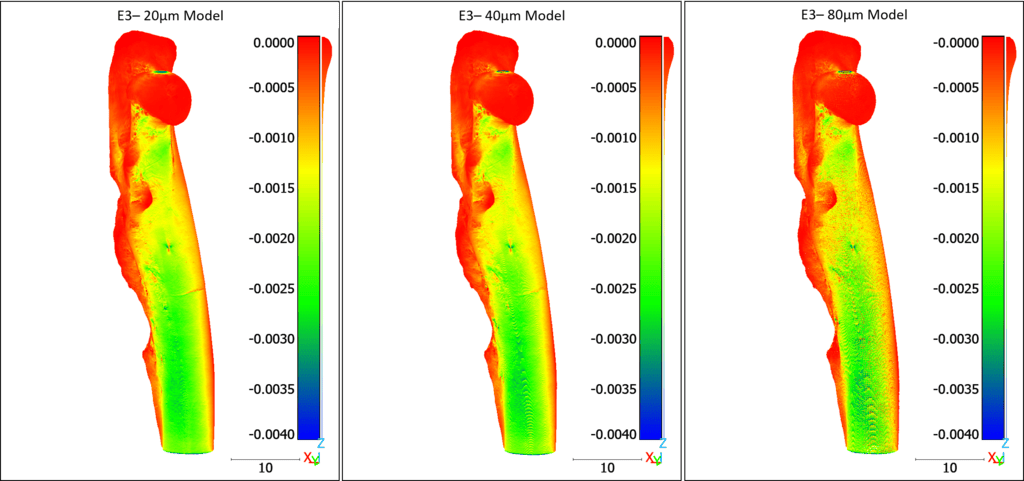}
        \caption{E3, Minimum principal strains.}
        \label{fig:E320vs40vs80}
    \end{subfigure}
    \caption{Visual comparison of the displacement magnitude (Umag) and principal strain fields (\text{E1} and \text{E3}) for the high-resolution 20 $\mu$\text{m},  coarsened 40 $\mu$\text{m}, and 80 $\mu$\text{m} MFEM models. After applying identical post-processing, the tree resolutions exhibit nearly identical global displacement patterns and Gaussian-smoothed strain distributions.}
    \label{fig:Results20vs40vs80}
\end{figure}

\begin{figure}[H]
    \centering
    \begin{subfigure}[b]{0.49\textwidth}
        \centering
        \includegraphics[height=5cm]{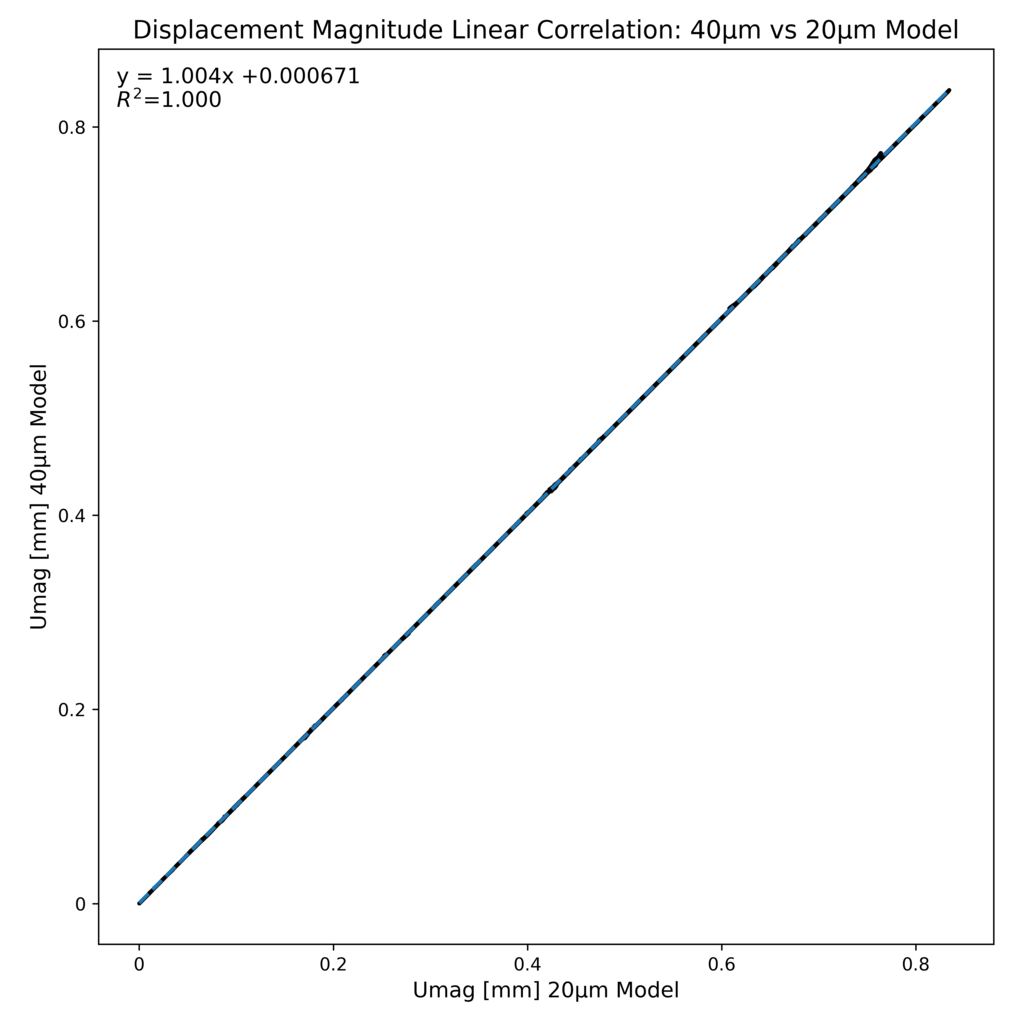}
        \caption{20~$\mu$m vs.\ 40~$\mu$m, \text{8,927,923 / 9,115,760} neighbor data points were compared.}
        \label{fig:LC20vs40Umag}
    \end{subfigure}
    \hfill
    \begin{subfigure}[b]{0.49\textwidth}
        \centering
        \includegraphics[height=5cm]{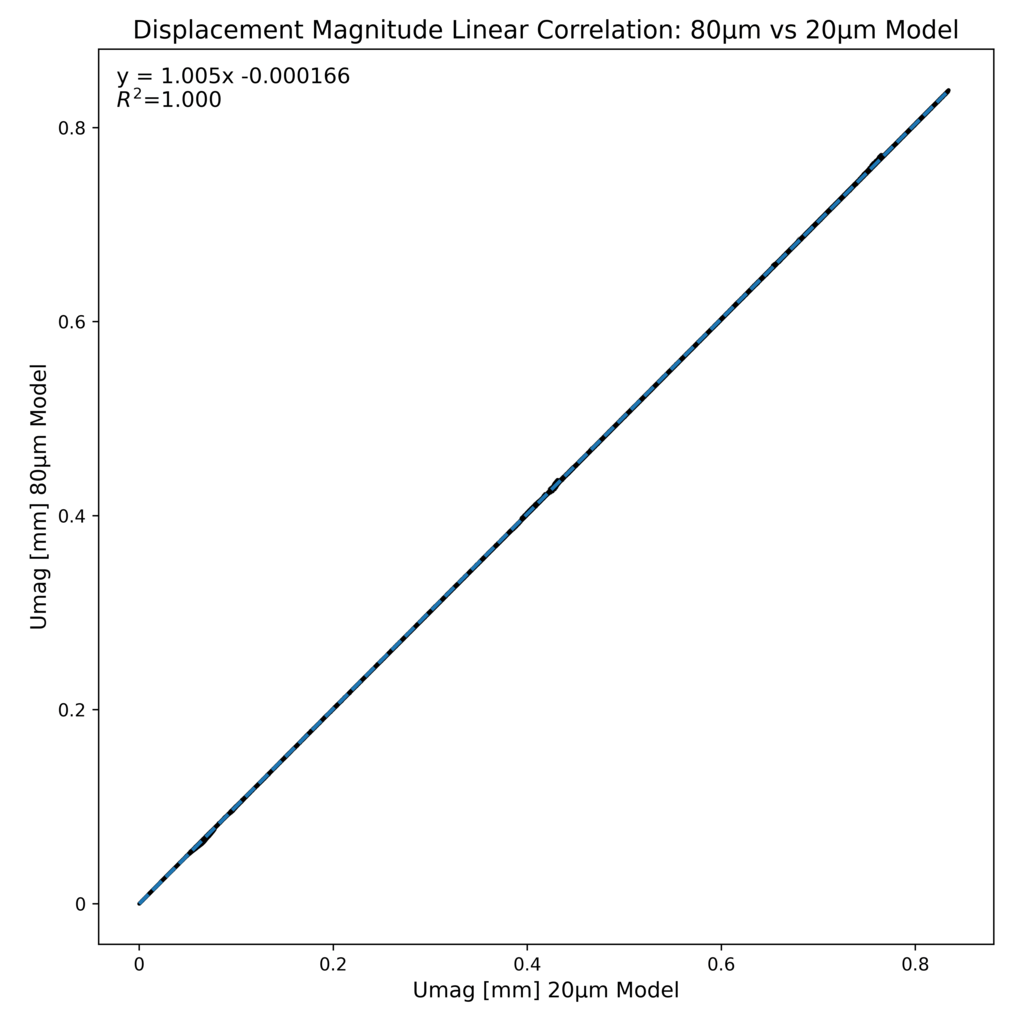}
        \caption{20~$\mu$m vs.\ 80~$\mu$m, \text{1,969,503 / 1,972,701} neighbor data points were compared.}
        \label{fig:LC20vs80Umag}
    \end{subfigure}
    \caption{Displacement magnitude [mm] comparison between the MFEM's L20 models at three resolutions.}
    \label{fig:LC20vs40vs80Umag}
\end{figure}

\begin{figure}[H]
    \centering
    \begin{subfigure}[b]{0.49\textwidth}
        \centering
        \includegraphics[height=5cm]{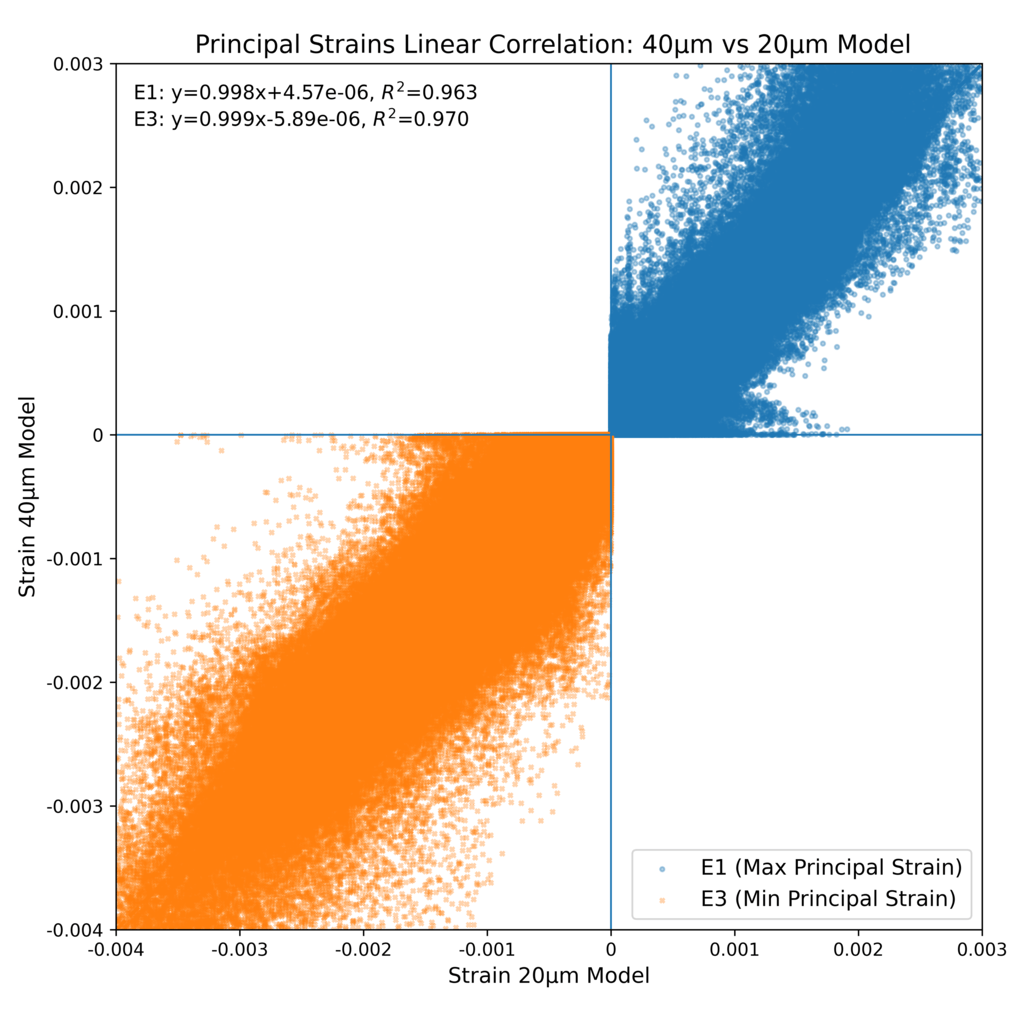}
        \caption{20~$\mu$m vs.\ 40~$\mu$m, \text{8,927,923 / 9,115,760} neighbor data points were compared.}
        \label{fig:LC20vs40E1E3}
    \end{subfigure}
    \hfill
    \begin{subfigure}[b]{0.49\textwidth}
        \centering
        \includegraphics[height=5cm]{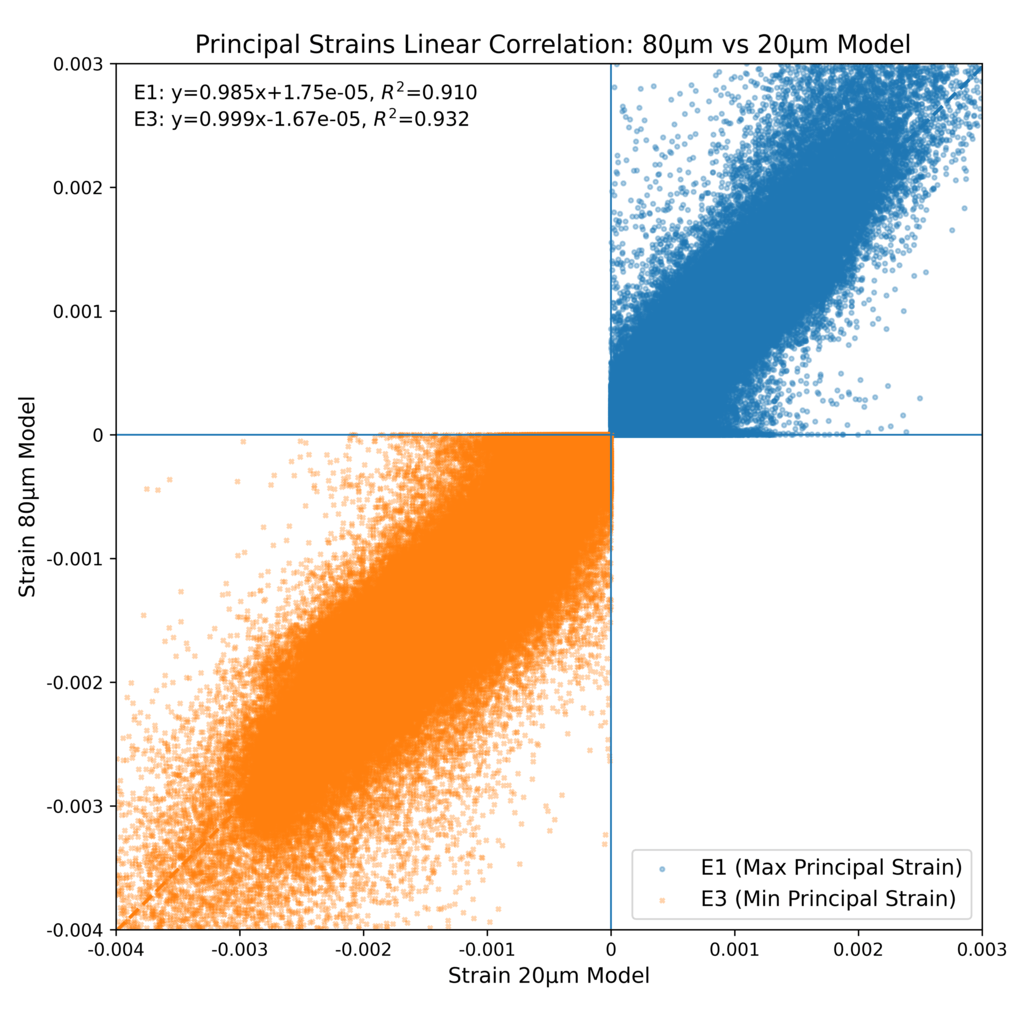}
        \caption{20~$\mu$m vs.\ 80~$\mu$m, \text{1,969,503 / 1,972,701} neighbor data points were compared.}
        \label{fig:LC20vs80E1E3}
    \end{subfigure}
    \caption{Principal strains E1 and E3 comparison between the MFEM's L20 models at three resolutions.}
    \label{fig:LC20vs40vs80E1E3}
\end{figure}

To assess the effect of voxel-size coarsening on the predicted displacement and strain fields, we compared the original 20~$\mu$m voxel model (L20) with two downsampled models of 40~$\mu$m and 80~$\mu$m voxel sizes. The raw $\mu$CT volume was downsampled in Simpleware, so that 2×2×2 native voxels were merged into a single 40~$\mu$m voxel, and each block of 4×4×4 voxels into a single 80~$\mu$m voxel. The downsampled datasets were segmented using the MIA clustering algorithm and converted into $\mu$FE models in Simpleware.
All models were solved by MFEM under identical boundary conditions: a 91~N compressive load applied to the femoral head, and the shaft distal surface was fully clamped. Following the FE solution, only the outer-boundary nodes were retained, and the resulting surfaces were rigidly aligned in CloudCompare for point-wise comparison. Strain fields were smoothed using a Gaussian filter (radius $r = 0.1$~mm, $\sigma = r/2$) to obtain a continuous boundary strain field. Correspondence between models was established by a nearest-neighbour search: a radius of 0.04~mm yielded 8,927,923 / 9,115,760 matched points for the 20~$\mu$m vs.\ 40~$\mu$m comparison, and 0.08~mm yielded 1,969,503 / 1,972,701 matched points for the 20~$\mu$m vs.\ 80~$\mu$m comparison.
Figure \ref{fig:Results20vs40vs80} summarizes the displacement and principal strain correlations for all models. Both the 40~$\mu$m and 80~$\mu$m models reproduced the displacement magnitude of the 20~$\mu$m reference almost perfectly. Linear regression for the 40~$\mu$m model yielded
\begin{align*}
\text U_{mag}^{40} = 1.004\,U_{mag}^{20} + 6.7\times10^{-4}\quad (R^2=1.000),
\end{align*}
corresponding to a negligible slope bias of \text{+0.4\%} (Fig.~\ref{fig:LC20vs40Umag}). The 80 $\mu$m model showed similarly excellent agreement
\begin{align*}
\text U_{mag}^{80} = 1.005\,U_{mag}^{20} - 1.66\times10^{-4}\quad (R^2=1.000),
\end{align*}
indicating a slop bias of \text{+0.5\%} (Fig.~\ref{fig:LC20vs80Umag}). Thus, global displacement predictions were essentially invariant across the 20--80 $\mu$m range.
For the Gaussian-smoothed principal strains, the 40 $\mu$m model exhibited excellent agreement with the 20 $\mu$m reference for both strain components (see Fig. ~\ref{fig:LC20vs40E1E3})
\begin{align*}
\text E1^{40} = 0.998\,E1^{20} + 4.6\times10^{-6}\quad (R^2=0.963), \\
\text E3^{40} = 0.999\,E3^{20} - 5.9\times10^{-6}\quad (R^2=0.970).
\end{align*}
These slopes correspond to biases of \text{-0.2\%} for \text{E1} and \text{-0.1\%} for \text{E3}, indicating that strain predictions at 40~$\mu$m are close to those at 20~$\mu$m.
The 80 $\mu$m model showed a slightly larger deviation (see Fig. ~\ref{fig:LC20vs80E1E3})
\begin{align*}
\text E1^{80} = 0.985\,E1^{20} + 1.75\times10^{-5}\quad (R^2=0.910), \\
\text E3^{80} = 0.999\,E3^{20} - 1.67\times10^{-5}\quad (R^2=0.932).
\end{align*}
corresponding to a modest slope bias of \text{-1.5\%} for \text{E1} and \text{-0.1\%} for \text{E3}, however a much larger spread in strains.

In general, coarsening the $\mu$CT-derived $\mu$FE model from 20~$\mu$m to 40~$\mu$m preserved boundary displacements but had some influence (although not dramatic) on principal strain predictions with $<1\%$ slope bias. Coarsening to 80~$\mu$m increased the variability between corresponding points, as reflected by the much lower $R^2$ values. These results indicate that a voxel size of 40~$\mu$m is sufficient for good prediction of boundary-level displacements and principal strains in the NZW rabbit femur, while 80~$\mu$m remains suitable for capturing global behavior, but shows reduced fidelity in reproducing finer spatial variations of the strain field.

\section{Sensitivity Analyses and Validation by Experimental Observations} \label{s.Sensitivity}

\subsection{\texorpdfstring{%
Comparison of L40 $\mu$FE Models Generated with Different MIA Grid Sizes}{%
Comparison of L40 µFE Models Generated with Different MIA Grid Sizes}}

The MIA clustering algorithm \cite{Dunmore2018MIA-Clustering:Material} is a machine learning-based approach that combines two clustering algorithms with minimal user input. Initially, K-means clustering \cite{Lloyd1982LeastPCM} segments $\mu$CT scans based on voxel gray values, followed by a fuzzy c-means algorithm \cite{PhamAnInhomogeneities}, which iteratively estimates membership probabilities, and each voxel is assigned to the structure corresponding to its highest membership probability.
In cases of gray value inhomogeneity, global clustering may lose fine details. To address this, a secondary clustering step, local fuzzy c-means is applied. The volume is subdivided into overlapping cubes using a user-defined grid size parameter (in voxel units). For each cube, local cluster membership probabilities are summed, and clusters falling below a threshold (default: 0.02) are ignored \cite{Dunmore2018MIA-Clustering:Material}. The final segmentation is obtained by merging overlapping cluster probabilities and assigning each voxel based on its highest membership probability.
To balance fine detail segmentation and larger structures, the grid size parameter should be slightly larger than the largest dimension of the region of interest, such as the thickest trabecula \cite{Dunmore2018MIA-Clustering:Material}.
Variations in grid size value affect the $\mu$FE results, thus we investigate the sensitivity of the value on the FE results.

\begin{figure}[H] \centering
    \includegraphics[width={1.0\textwidth}]{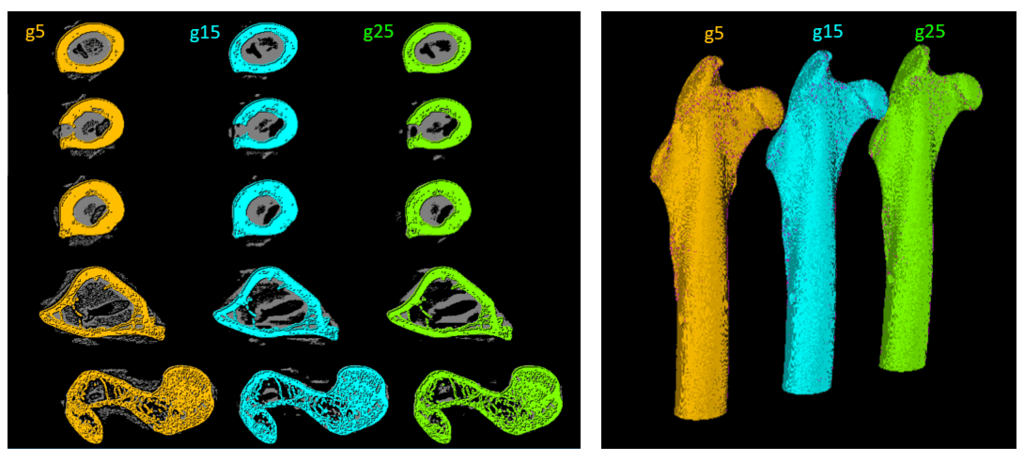}
    \caption{MIA's clustering algorithm raw segmented output with various grid sizes of 5, 15, and 25 voxels for L40 $\mu$CT scans.  Left: 2D slices. Right: 3D geometry.}
    \label{fig:MIArawOutput}
\end{figure}

The L40 $\mu$CT scan with a resolution of 40 $\mu$m was used as the basis of the sensitivity investigation. The thickest trabecula in the neck region, representing the region of interest, measures approximately 0.4 mm. This corresponds to a standard grid size of 15 voxels, with 0.4 mm covering a central region of 10 voxels, plus an additional margin of 2.5 voxels on each side. To investigate the impact of grid size, three distinct sizes were applied to MIA segmentation process: 5, 15, and 25 voxels.
Fig.\ref{fig:MIArawOutput} shows the output for the three segmentations, illustrating grid size influences on detail level. Smaller grid sizes capture more details within the proximal section of the femur, containing trabecular structure, while fewer details are resolved in the dense cortical bone of the shaft distal area. Simpleware was used to create the $\mu$FE models from the segmented scans by removing floating voxels from the main bone body, defining the boundary surface, and generating the mesh.
The applied traction force was carefully calibrated to each $\mu$FE mesh file, ensuring a 100 N vertical load applied to the femoral head, while the distal femur surface was constrained.  Table  \ref{tab:MIAelement} presents the number of bone elements for each $\mu$FE model, referred to as L40g5, L40g15, and L40g25.
\begin{table}[ht]
    \begin{center}
    \begin{tabular}{|l|c|c|c|}
        \hline
        Mesh & \# of Elements & DOFs \\
        \hline
        L40g5 & 36,717,664 & 126,004,782 \\
        \hline
        L40g15 & 36,994,360 & 125,246,766 \\
        \hline
        L40g25 & 37,555,013 & 126,149,589 \\
        \hline
    \end{tabular}
    \caption{Number of elements for each $\mu$FE model constructed by MIA segmentation with different grid size input after removal of unconnected islands in Simpleware.}
    \label{tab:MIAelement}
\end{center}
\end{table}

\begin{figure}[H]
    \centering
    \begin{subfigure}[b]{0.49\textwidth}
        \includegraphics[width=\linewidth]{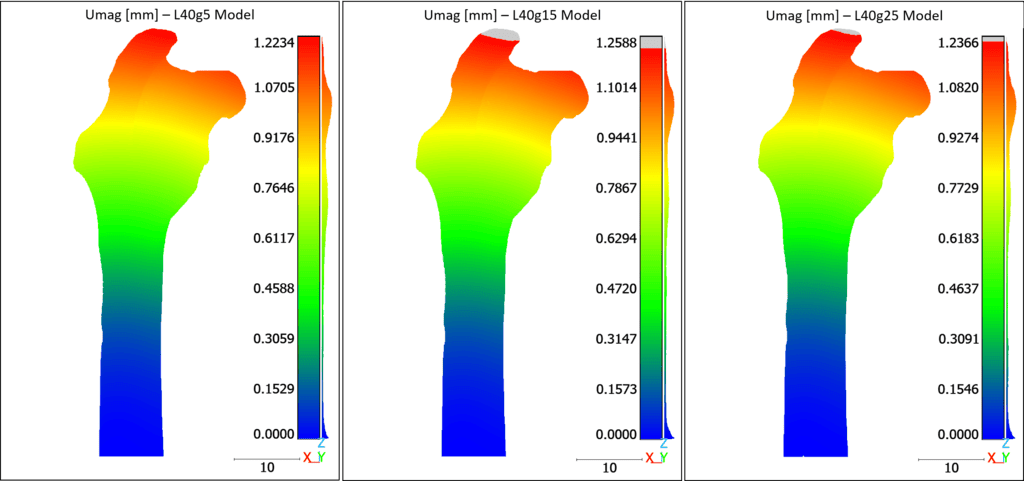}
        \caption{Umag, Displacement magnitude [mm].}
        \label{fig:Umag5vs15vs25}
    \end{subfigure}
    \vspace{1em}

    \begin{subfigure}[b]{0.49\textwidth}
        \includegraphics[width=\linewidth]{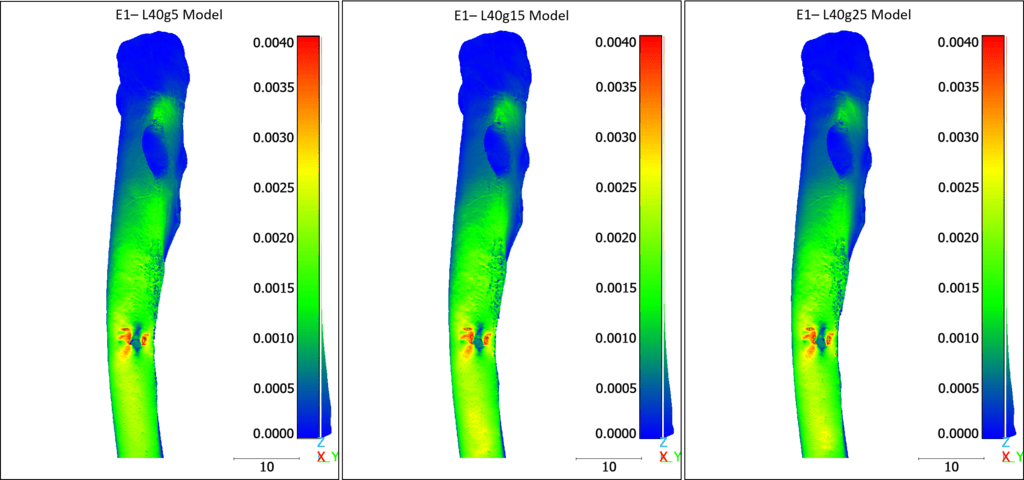}
        \caption{E1, Maximum principal strains.}
        \label{fig:E15vs15vs25}
    \end{subfigure}
    \hfill
    \begin{subfigure}[b]{0.49\textwidth}
        \includegraphics[width=\linewidth]{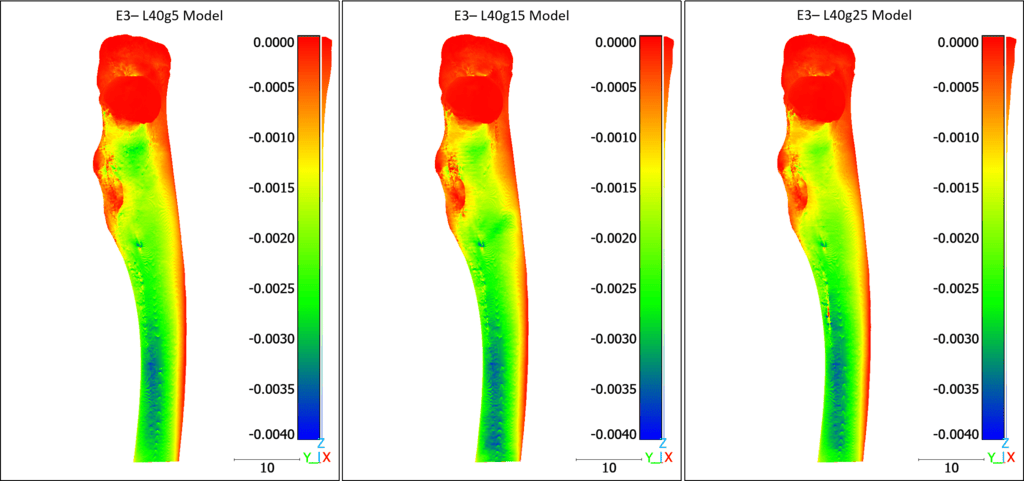}
        \caption{E3, Minimum principal strains.}
        \label{fig:E35vs15vs25}
    \end{subfigure}
    \caption{Visual comparison of the displacement magnitude (Umag) and principal strain fields (\text{E1} and \text{E3}) for L40 model (40 $\mu$\text{m}) with different segmentation grid size: 5 (L40g5), 15 (L40g15), and 25 (L40g25) voxels. After applying identical post-processing, the three segmentations exhibit nearly identical global displacement patterns and Gaussian-smoothed strain distributions.}
    \label{fig:MIA_Results5vs15vs25}
\end{figure}

\begin{figure}[H]
    \centering
    \begin{subfigure}[b]{0.49\textwidth}
        \centering
        \includegraphics[height=5cm]{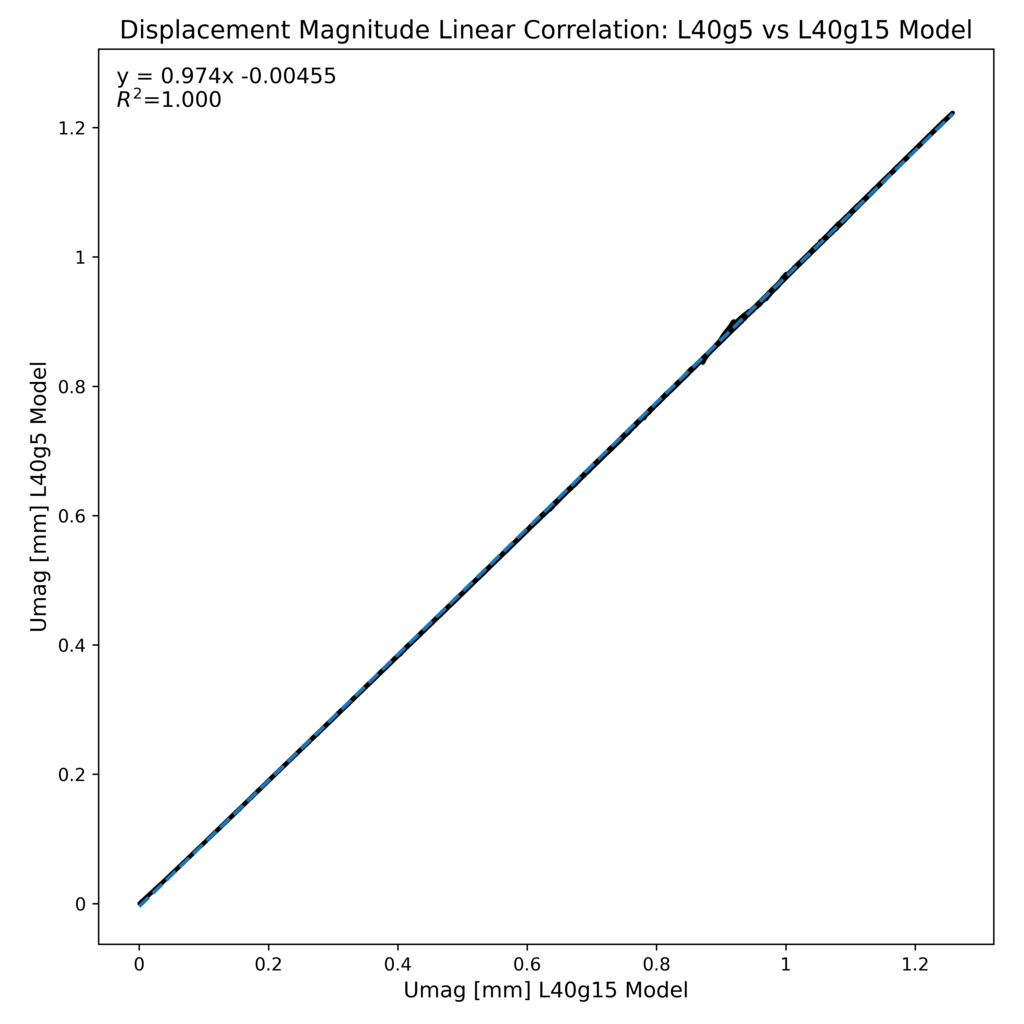}
        \caption{ L40g5 vs. L40g15, \text{8,142,302 / 11,094,022} neighbor data points were compared.}
        \label{fig:LCg15vsg5Umag}
    \end{subfigure}
    \hfill
    \begin{subfigure}[b]{0.49\textwidth}
        \centering
        \includegraphics[height=5cm]{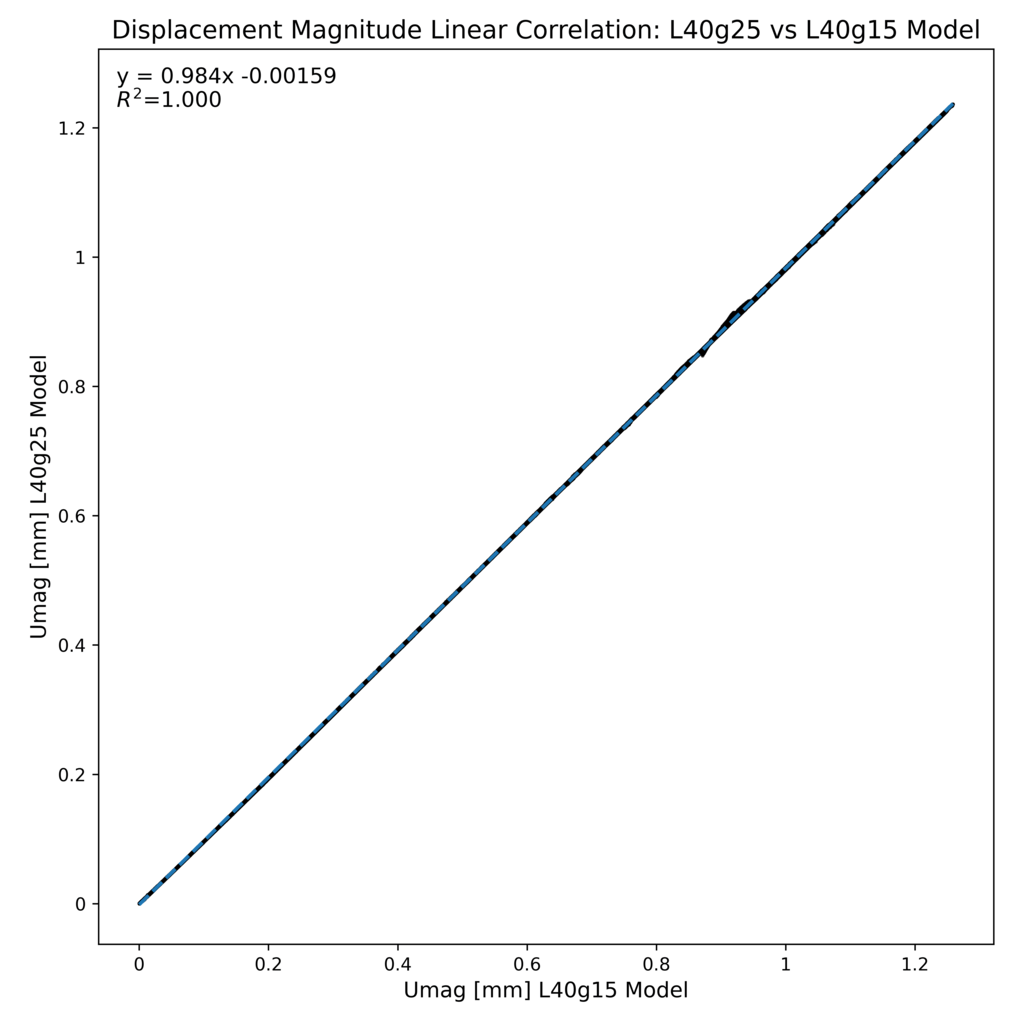}
        \caption{L40g25 vs. L40g15, \text{8,213,770 / 9,238,825} neighbor data points were compared.}
        \label{fig:LCg15vsg25Umag}
    \end{subfigure}
    \caption{Displacement magnitude [mm] comparison between the MFEM's L40 models at three MIA grid sizes.}
    \label{fig:LCg15vsg5vsg25Umag}
\end{figure}

\begin{figure}[H]
    \centering
    \begin{subfigure}[b]{0.49\textwidth}
        \centering
        \includegraphics[height=5cm]{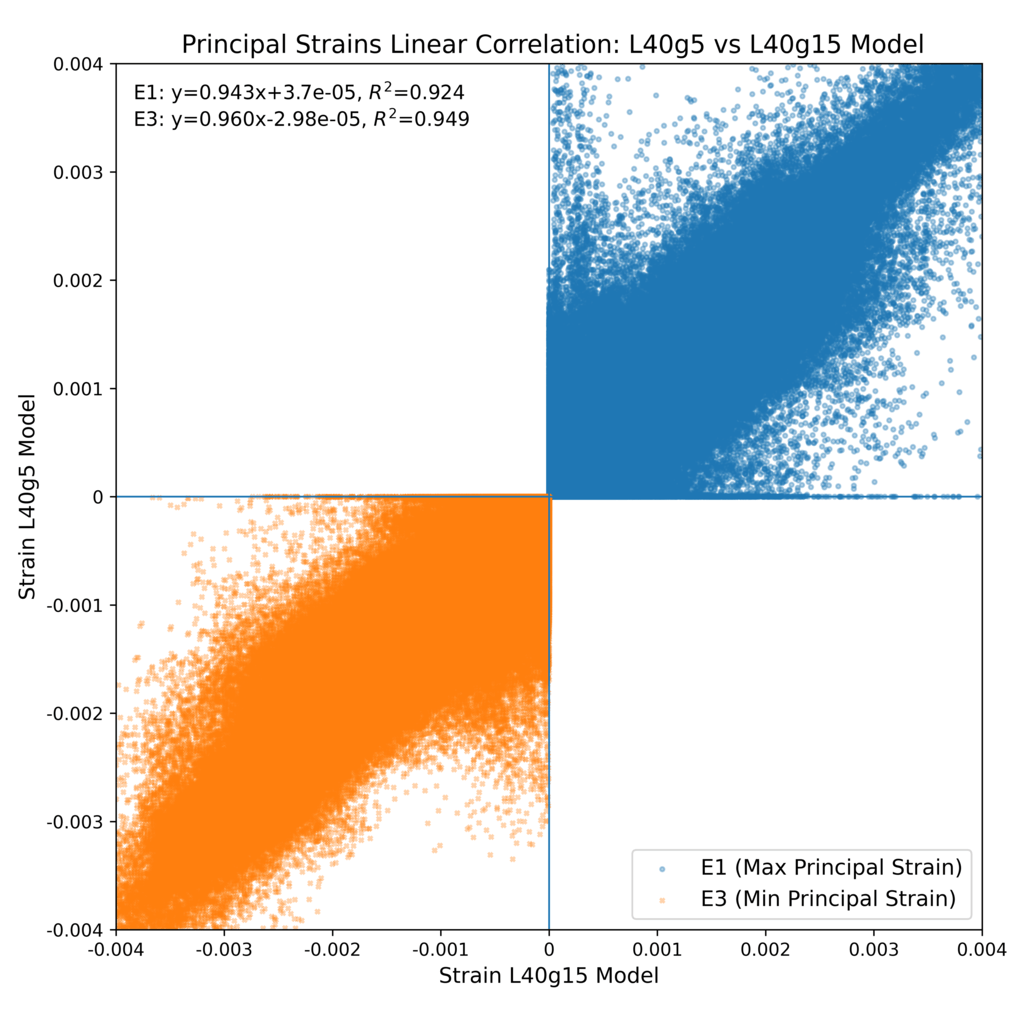}
        \caption{L40g5 vs. L40g15, \text{8,142,302 / 11,094,022} neighbor data points were compared.}
        \label{fig:LCg15vsg5E1E3}
    \end{subfigure}
    \hfill
    \begin{subfigure}[b]{0.49\textwidth}
        \centering
        \includegraphics[height=5cm]{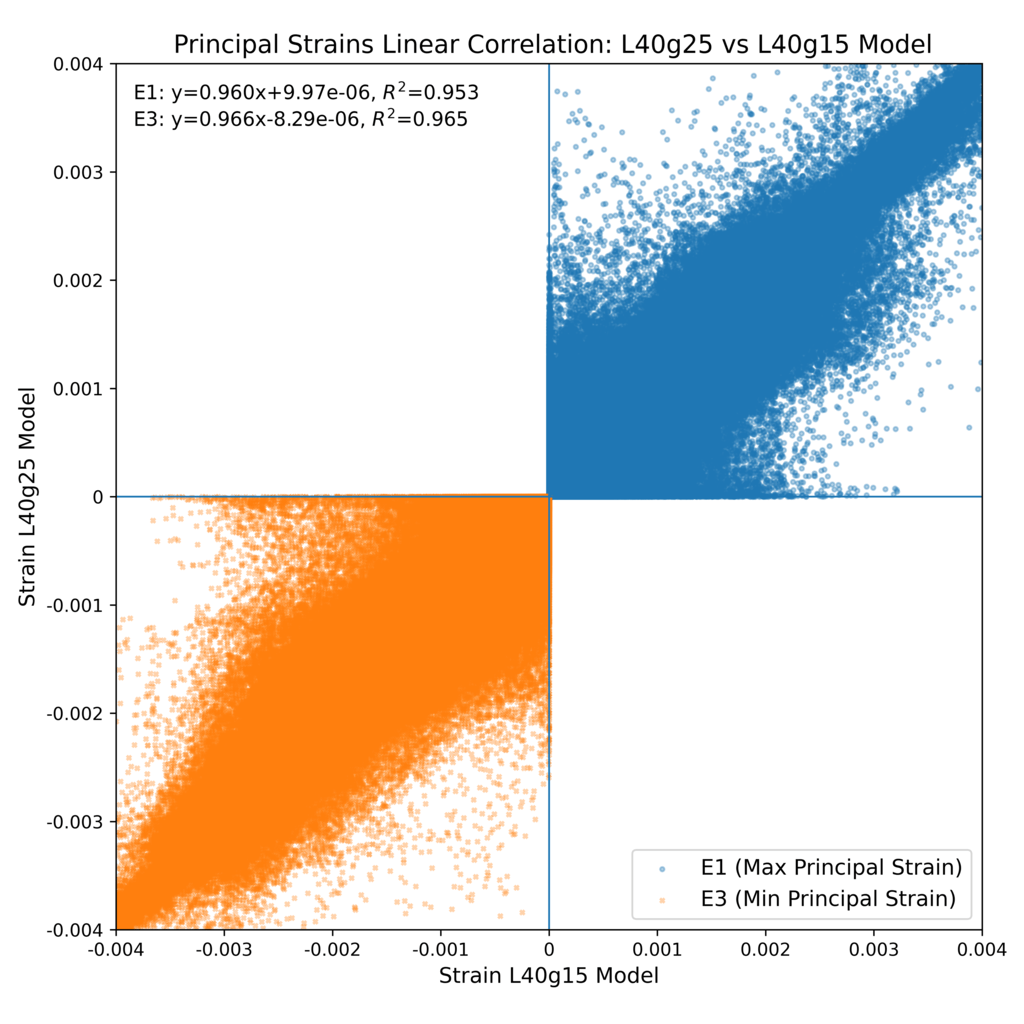}
        \caption{L40g25 vs. L40g15, \text{8,213,770 / 9,238,825} neighbor data points were compared.}
        \label{fig:LCg15vsg25E1E3}
    \end{subfigure}
    \caption{Principal strains E1 and E3 comparison between the MFEM's L40 models at three MIA grid sizes.}
    \label{fig:LCg15vsg5vsg25E1E3}
\end{figure}

The resulting $\mu$FE models were solved using MFEM.
Outer-boundary nodes were extracted, and the principal strains were smoothed using a Gaussian filter with radius
$r = 0.16~\text{mm}$ ($\approx 4$ voxels, $\sigma = r/2$).
Figure~\ref{fig:MIA_Results5vs15vs25} shows the displacement magnitude and principal strain distributions for all models.
The boundary surfaces were rigidly registered in CloudCompare and paired using a nearest-neighbor search with a 0.065~mm radius, yielding
8,142,302 / 11,094,022 matched points for L40g15 vs.\ L40g5 and
8,213,770 / 9,238,825 matched points for L40g15 vs.\ L40g25.
The displacement magnitude ($U_{\mathrm{mag}}$) was highly consistent across all three models, with negligible global differences.
Linear regression between L40g5 and L40g15 produced
\begin{align*}
\text U_{mag}^{g5} = 0.974\,U_{mag}^{g15} - 4.5\times10^{-3}\quad (R^2=1.000),
\end{align*}
and between L40g25 and L40g15
\begin{align*}
\text U_{mag}^{g25} = 0.984\,U_{mag}^{g15} - 1.6\times10^{-3}\quad (R^2=1.000).
\end{align*}
Both slopes indicate a bias of less than $-4\%$ (Fig.~\ref{fig:LCg15vsg5vsg25Umag}), confirming that the global stiffness and overall displacement response are essentially insensitive to the segmentation grid size.
In contrast, the principal strains exhibited higher sensitivity to the grid size parameter.
After Gaussian smoothing ($r=0.16$~mm), the linear relationships for L40g5 relative to L40g15 (Fig.~\ref{fig:LCg15vsg5E1E3}) were
\begin{align*}
\text E1^{g5} = 0.943\,E1^{g15} + 3.7\times10^{-5}\quad (R^2=0.924), \\
\text E3^{g5} = 0.960\,E3^{g15} - 2.98\times10^{-5}\quad (R^2=0.949),
\end{align*}
corresponding to biases of $-6\%$ for \text{E1} and $-4\%$ for \text{E3}.
For L40g25 relative to L40g15 (Fig.~\ref{fig:LCg15vsg25E1E3})
\begin{align*}
\text E1^{g25} = 0.960\,E1^{g15} + 9.97\times10^{-6}\quad (R^2=0.953), \\
\text E3^{g25} = 0.966\,E3^{g15} - 8.29\times10^{-6}\quad (R^2=0.965),
\end{align*}
corresponding to biases of $-4\%$ for \text{E1} and $-3\%$ for \text{E3}.
Varying the MIA grid size between 5--25 voxels had negligible influence on the global mechanical response (displacements) and produced a modest, systematic deviations in the principal strains. The larger strain scatter observed for L40g5 reflects increased trabecular detail, whereas L40g25 yields slightly smoother and slightly higher strain amplitudes. Overall, a grid size of 15 voxels offers a good balance: it preserves mechanically relevant strain patterns while avoiding unnecessary segmentation noise, and alternative grid sizes remain within a predictable bias band.

\subsection{Determining NZW Rabbit's Bone Properties}

Finally, we conclude by presenting the methodology for calibrating bone material properties at the micron scale by comparing experimental observation using DIC with the L40 $\mu$FE model. A detailed description of the experimental campaign and biomechanical analysis is discussed in a different publication under review.

\begin{figure}[H]
    \centering
    \begin{subfigure}[b]{0.49\textwidth}
        \includegraphics[width=\linewidth]{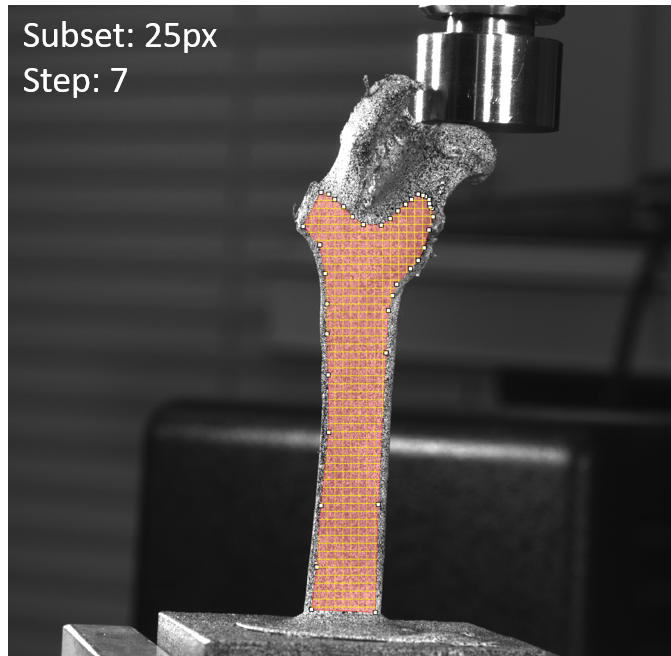}
        \caption{L40 DIC ROI1.}
        \label{fig:L40test3ROIsys1}
    \end{subfigure}
    \hfill
    \begin{subfigure}[b]{0.49\textwidth}
        \includegraphics[width=\linewidth]{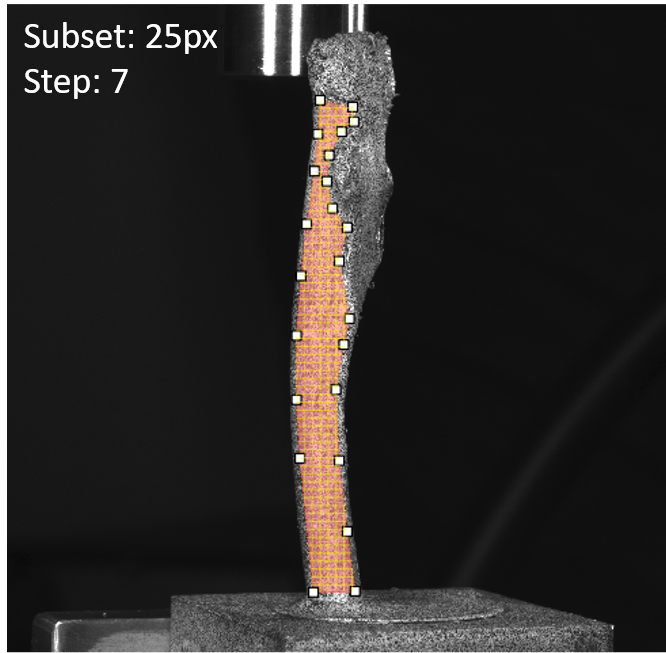}
        \caption{L40 DIC ROI2.}
        \label{fig:L40test3ROIsys}
    \end{subfigure}
    \caption{The L40 NZW Rabbit femur, two ROI's monitored by the DIC technique.}
    \label{fig:L40DIC_ROI}
\end{figure}

\begin{figure}[H]
    \centering
    \includegraphics[width=\linewidth]{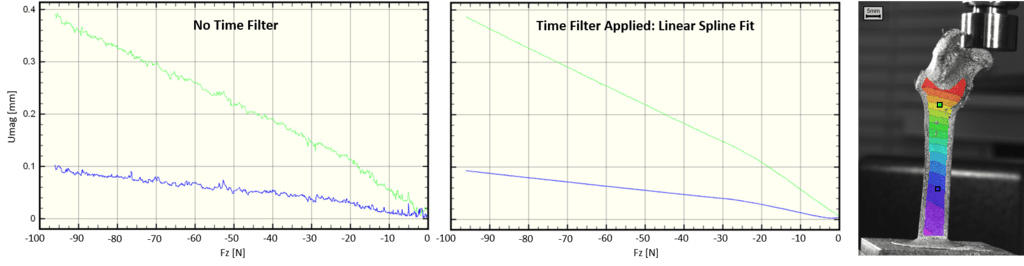}
    \caption{ L40 DIC ROI1:
            \textbf{Left:} Displacement magnitude versus force ($F_z$) at two selected points in the DIC-measured field, showing the raw displacement noise over time.
            \textbf{Center:} The same data after applying a time filter using a linear spline fit, which effectively smooths the random displacement fluctuations.
            \textbf{Right:} The full-field displacement magnitude map obtained from the DIC measurement at a certain time.
            }
    \label{fig:L40test3Umag_sys1}
\end{figure}

\begin{figure}[H]
    \centering
    \includegraphics[width=\linewidth]{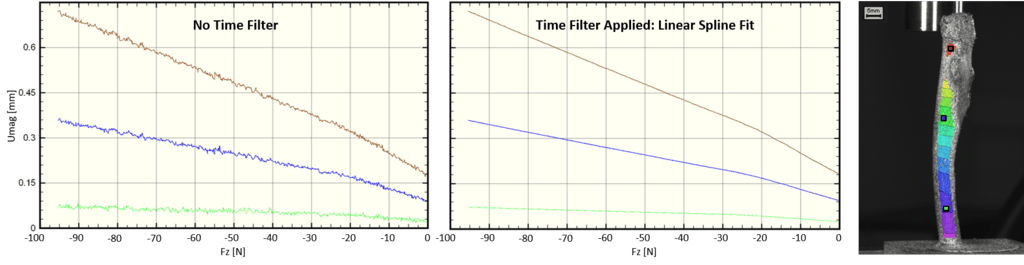}
    \caption{ L40 DIC ROI2:
            \textbf{Left:} Displacement magnitude versus force ($F_z$) at two selected points in the DIC-measured field, showing the raw displacement noise over time.
            \textbf{Center:} The same data after applying a time filter using a linear spline fit, which effectively smooths the random displacement fluctuations.
            \textbf{Right:} The full-field displacement magnitude map obtained from the DIC measurement at a certain time.
            }
    \label{fig:L40test3Umag_sys}
\end{figure}

\begin{figure}[H]
    \centering
    \begin{subfigure}[b]{0.4\textwidth}
        \includegraphics[width=\linewidth]{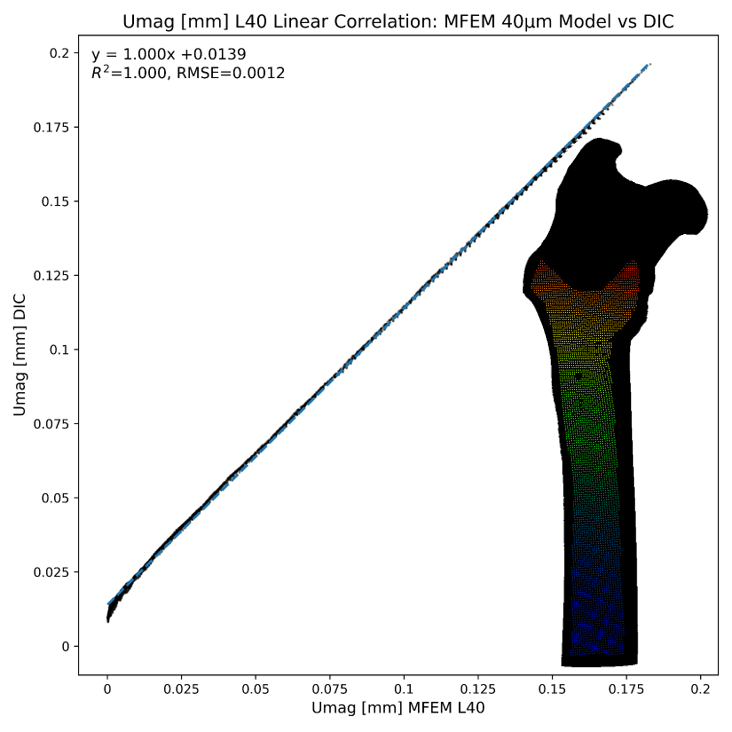}
        \caption{ROI1 linear correlation.}
        \label{fig:L40vsDIC_Umag_sys1}
    \end{subfigure}
    \hfill
    \begin{subfigure}[b]{0.4\textwidth}
        \includegraphics[width=\linewidth]{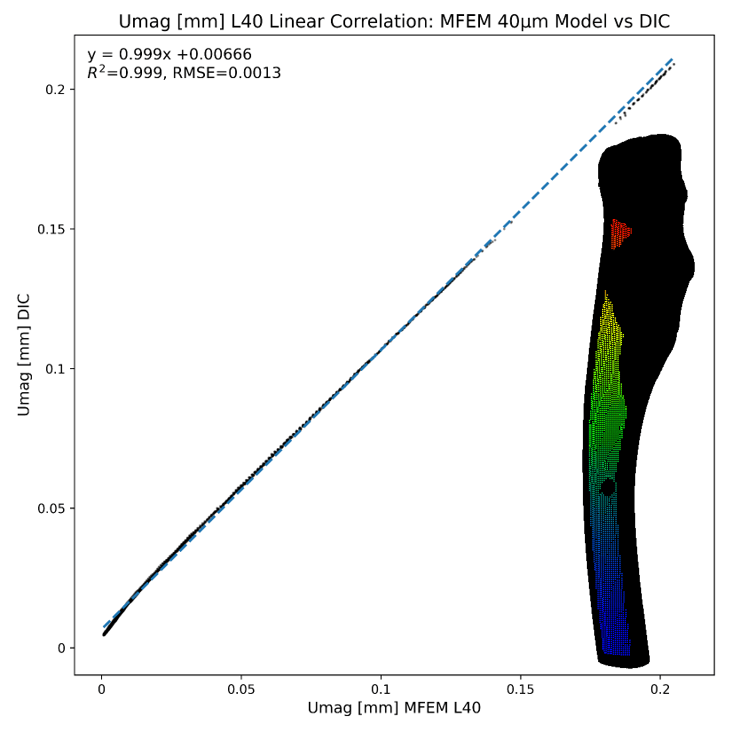}
        \caption{ROI2 linear correlation.}
        \label{fig:L40vsDIC_Umag_sys}
    \end{subfigure}
    \caption{Comparison between DIC and L40~$\mu$FE model displacement magnitudes after rigid-body correction and registration in CloudCompare. The $\mu$FE model displacements were scaled by~0.47 to match the experimental response at both ROI's, equivalent to a corrected Young's modulus $\sim$2.1~times the assumed 10~GPa. A total of 3{,}478 in ROI1 and 2{,}428 in ROI2 matched points show an almost perfect correlation ($R^{2}\!\approx\!1.0$, RMSE$=0.00125$).}
    \label{fig:L40vsDIC_Umag}
\end{figure}

Two independent Digital Image Correlation (DIC) systems were employed to monitor the mechanical response of the rabbit femur at two regions of interest (ROIs), as illustrated in Fig.~\ref{fig:L40DIC_ROI}. The displacement magnitude field was computed for both ROIs, exhibiting the expected linear-elastic behavior (see Fig.~\ref{fig:L40test3Umag_sys1}-\ref{fig:L40test3Umag_sys}). To isolate the deformation field from rigid-body motion, two arbitrary loading steps were selected and their displacement-magnitude fields were subtracted, yielding a corrected field corresponding to a 39~N load. The processed DIC point cloud was registered to the L40 $\mu$FE model in CloudCompare for optimal spatial alignment. We compared displacement magnitudes of ROI1 3{,}478 matched points (out of 4{,}280 DIC points; matching radius = 0.05~mm, $k=8$) and ROI2 2{,}428 matched points (out of 2{,}602 DIC points; matching radius = 0.05~mm, $k=8$) using linear regression and the root-mean-square error (RMSE). Both comparisons showed an almost perfect linear correlation between the L40 $\mu$FE model and the DIC fields (slope $=0.47$, $R^{2}\!\approx\!1.0$), indicating that the DIC magnitudes are approximately $47\%$ of the model predictions. After scaling the model displacements by 0.47 (Fig.~\ref{fig:L40vsDIC_Umag}), the agreement becomes nearly one-to-one. This implies that the estimated Young's modulus is $2.1\times$ larger than the 10~GPa value assigned in the L40 $\mu$FE simulation.

 \section{Summary and Conclusions} \label{s.Summary}

This work presented a comprehensive voxel-based $\mu$FE framework for large-scale biomechanical analysis of bone, combining high-resolution imaging, advanced segmentation, scalable finite element solvers, and experimental validation.
First, a complete pipeline for generating and solving $\mu$FE models from $\mu$CT data was established. Following specimen preparation and scanning, bone segmentation was performed using the MIA clustering algorithm, enabling robust separation of bone and non-bone phases at the voxel level. The segmented volumes were converted into voxel-based $\mu$FE meshes using Simpleware and solved using MFEM, employing algorithms specifically designed for extremely large linear elastic systems.
A novel matrix-free voxel-based multigrid solver was developed for the solution of very large FE models without reduced storage requirements. The performance of the newly developed solver was compared with the black-box algebraic multigrid.  The results demonstrate that MFEM provides a flexible and scalable platform capable of handling hundreds of millions of degrees of freedom while maintaining numerical robustness and solver efficiency on HPC systems.

Second, the accuracy and consistency of the $\mu$FE solutions were systematically verified. A comparison between MFEM and the commercial finite element solver Abaqus/CAE v2022 for a largely coarsen model showed perfect agreement in displacement field and a minor discrepancy of approximately $2\%$ in principal strain fields, validating the MFEM formulation for large-scale  $\mu$FE bone models. In addition, the efficiency of two MFEM's numerical approaches: full assembly and element-by-element was assessed, highlighting the trade-offs between memory footprint and computational time for voxel-based models. A $\mu$FE bone model with 811 million DOFs was solved using 192 MPI processes and  ~700 GB of memory in less than 5~hours, demonstrating the feasibility of solving anatomically realistic models at this scale.
The influence of spatial resolution was investigated by comparing $\mu$FE models generated from the same bone sample at voxel sizes of 20~$\mu$m, 40~$\mu$m, and 80~$\mu$m. Coarsening from 20~$\mu$m to 40~$\mu$m preserved boundary displacement and principal strain distributions with minimal bias ($<1\%$), while substantially reducing computational cost. Although the 80~$\mu$m model captured the global mechanical response, it exhibited increased scatter in local principal strain fields. Complementary sensitivity analyses examined the impact of MIA segmentation grid size (5, 15, and 25 voxels), demonstrating that this user-defined parameter can influence the global mechanical response by approximately $4\%$ and should therefore be carefully selected in resolution-dependent studies.
For all numerical comparisons, a dedicated post-processing procedure was introduced, enabling consistent local comparison of displacement magnitudes and principal strains across different numerical formulations, mesh resolutions, and solvers by focusing on boundary-level quantities.
Finally, a methodology for determining effective bone material properties at the micro scale was presented by coupling $\mu$FE simulations with experimental measurements. Two independent DIC systems were used to measure displacement fields at two regions of interest on an NZW rabbit femur. After registering the experimental point clouds to the $\mu$FE models, a strong linear correlation was observed between experimentally measured and simulated displacement magnitudes, implying an effective Young's modulus of approximately 21~GPa. A detailed description of the experimental campaign and biomechanical analysis is provided in a companion publication currently under review.

In conclusion, this study demonstrates that MFEM is a powerful and reliable tool for large-scale biomechanical $\mu$FE modeling of bone. When combined with robust segmentation as MIA, efficient numerical formulations, and carefully designed experimental validation, MFEM enables scalable, reproducible, and quantitatively validated simulations. The proposed framework provides a solid foundation for future investigations of preclinical treatment-related risks, with the potential to reduce reliance on animal testing.

\section*{Acknowledgements}

SMW and ZY thank Prof.\ Gal Shafistein from Roswell Park Cancer Center in Buffalo, NY, USA and Dr.\ Roni Kolerman at School of Medicine at Tel-Aviv University for providing the NZW femurs.
WP acknowledges the support of NSF RTG DMS-2136228 and the computational resources provided by the Oregon Regional Computing Accelerator (Orca), funded by NSF CC*-2346732. ZY acknowledges the support of ISF-DFG (grant No.\ 626/25).

\printbibliography

\end{document}